\documentclass[aps,floatfix]{revtex4}
\usepackage{epsfig}
\usepackage{color}
\usepackage{amssymb}
\usepackage{amsmath}
\usepackage[english]{babel}
\usepackage{multirow}
\usepackage{graphicx}

\newcommand{\etal}{\it et al.}

\newcommand{\e}{{\rm e}}

\newcommand{\veo}{\varepsilon_0}

\begin{document}

\title{Transition Rates for a Rydberg Atom Surrounded by a Plasma}

\author{Chengliang Lin, Christian Gocke and Gerd R\"opke}
\affiliation{Universit\"at Rostock, Institut f\"ur Physik, 18051 Rostock, Germany}
\author{Heidi Reinholz}
\affiliation{Universit\"at Rostock, Institut f\"ur Physik, 18051 Rostock, Germany}
\affiliation{University of Western Australia School of Physics, WA 6009 Crawley, Australia}
\date{\today}

\begin{abstract}
We derive a quantum master equation for an atom coupled to a heat bath represented by a charged particle many-body
environment. In Born-Markov approximation, the influence of the plasma environment on the reduced
system is described by the dynamical structure factor. Expressions for the profiles of spectral lines are obtained.
Wave packets are introduced as robust states allowing for a quasi-classical description of Rydberg electrons. 
Transition rates for highly excited Rydberg levels are investigated. 
A circular-orbit wave packet approach has been applied, in order to describe the localization of electrons
within Rydberg states.
The calculated transition rates are in a good agreement with experimental data. 
\end{abstract}

\maketitle

PACS number(s): 03.65.Yz, 32.70.Jz, 32.80.Ee, 52.25.Tx

\section{Introduction}

Open quantum systems have been a fascinating area of research because
of its ability to describe the transition from
the microscopic to the macroscopic world. The appearance of the classicality
in a quantum system, i.e. the loss of quantum informations of a quantum
system can be described by decoherence resulting from the interaction of an
open quantum system with its surroundings{~\cite{Schlosshauer,JZKGKS03}}.

An interesting example for an open quantum system interacting with a plasma environment are highly excited
atoms, so-called Rydberg states, characterized by a large main quantum number.
Rydberg states play an important role in astrophysics to study stellar atmospheres {\cite{YNG07,VOS13}}.
Particularly, ionization processes of Rydberg states of hydrogen and helium and 
their recombination processes are significant for hydrogen and helium plasmas
in a very low-density environment which exists in stellar atmospheres
with weakly ionized layers {\cite{YNG07,AAM96,AAM97}}. Because the interaction with the 
plasma, characterized by the plasma frequency, is no longer small compared to the energy 
differences of quantum eigenstates, the surrounding plasma cannot be considered as a weak perturbation
of the excited atom. The time evolution, in particular transition rates, is modified as 
shown in this work. An essential problem is the construction of optimum, robust states.

Note that Rydberg states are energetically near to the continuum of scattering states.
The screening of a given ion by the free electrons and
neighboring ions in a plasma results in the reduction of the ionization potential and line broadening 
of eigenenergy levels of the given atom. For the Rydberg states near the continuum edge,
it may be quite difficult to rigorously distinguish the borderline between the real continuum
edge and bound states. For example, it is known that in solar astrophysics spectral lines are 
visible up to main quantum numbers of about 17 \cite{EKKR85}. 
The correct treatment of the Rydberg states which are near the continuum edge 
is a long-standing problem in plasma spectroscopy, see Refs.{~\cite{BO07,BO11}}. Thus
a many-body approach to Rydberg states in a plasma is also of interest for spectroscopy.

Because of their macroscopic characters and long lifetimes, nowadays
Rydberg states  
become a fundamental concept of open quantum systems in different fields of physics,
such as quantum information research{~\cite{SWM10,LRBBN14,RLBBL14}} and 
ultracold plasmas{~\cite{VCTP05,KPPR07,RS13}}.
Recently the existence of Rydberg excitons in the copper oxide Cu$_2$O{~\cite{KFSSB14}} is demonstrated 
which enable visible measurements of coherent quantum effects{~\cite{GAFBSS14}}.
Actually, using a localized semi-classical representation of bound states to study
the connection between classical mechanics and the large-quantum number 
limit of quantum mechanics has been a topic of interest since 
the development of quantum mechanics{~\cite{Sch26,Darwin28}}. 
As a mesoscopic object, the Rydberg atom may be regarded as an
outstanding example demonstrating both macroscopic classical and
microscopic quantum behavior. In a series of papers of Stroud {\etal}
{\cite{PS86,YMPS89,YMS90,YS91,MS94,NS96,PB98,BS99}},
the dynamics of a hydrogenic Rydberg atom has been discussed in detail.
It has been shown that the behavior of
a wave packet constructed by energy eigenstates of the 
hydrogen atom is different for the short and the long-term evolution.
This difference is essential for the investigation of 
the connection between the quantum and classical description of nature and gives 
a possibility to explain the emergence of classicality in a quantum
system.

Motivated by these exciting perspectives, we study the properties 
of hydrogenic Rydberg atoms, in particular, the transition rates of highly excited Rydberg states.
Different environments are of interest: the interaction with the radiation field,
the interaction with phonons (Rydberg excitons), the interaction with charged particles.
We focus on the special case where the environment is described by a plasma background,
see Refs. {\cite{GR06,GR07,GR08}}. A similar derivation
for a test particle interacting through collisions with a low-density background gas by 
using the quantum master equation approach is reported in Refs.{~\cite{VH09,SV10}}.
The influence of the plasma on the dynamics of the atom is determined by the dynamical structure 
factor of the surrounding plasma. Robust states are represented by optimized Gaussian wave packets.
As an example, transition rates are calculated and compared to 
other theoretical approaches and experimental data.

Another example which will be considered are the profiles of spectral lines.
They are essentially determined by the interaction
of the bound states with the radiation field and the charge carriers of the plasma. Both of them 
can be treated as thermal bath for the bound states, which are regarded as the reduced system in the
theory of open quantum systems. Various approaches can be used to calculate 
the spectral line profiles in a plasma environment, for instance, unified theory{~\cite{SC69}}, quantum
mechanical scattering theory{~\cite{Ba58}} and the Green's function methods{~\cite{Kl69,GHR91}}, which are 
based on the assumption that the plasma is in equilibrium. Quantum kinetic theory, as a nonequilibrium approach,
can also be applied to investigate the line profiles of the plasma which will be presented in this work.

This paper is organized as follows: in section{~\ref{sec:geqme:deriv}} we outline
the derivation of the general quantum master equation in Born-Markov 
approximation. Then we discuss the special case of plasma as a many-body environment in
Sec.{~\ref{sec:geqme:conn}}.
In Sec.{~\ref{sec:geqme:atom}},
the general quantum master equation is investigated in detail by introducing the basis of
the energy eigenstates of the hydrogen atom. 
The Pauli equation and the spectral line profiles are derived in this section.
The wave packet description for the bound Rydberg electron is introduced in Sec.{~\ref{sec:wavepa}}.
The robustness and validity of the wave packet description are discussed in Sec.{~\ref{wp:robust}}.
The transition rates for the hydrogenic Rydberg atom
derived with the use of the circular-orbit wave packet and their comparisons with 
classical Monte-Carlo simulations and experimental data are presented in{~\ref{wp:transrate}}.
Conclusions are drawn in Sec.~\ref{sec:Discussion}.

\section{Quantum Master Equation for Rydberg Atoms in a Plasma} 
\label{sec:geqme}

\subsection{General quantum master equation}
\label{sec:geqme:deriv}

We are investigating the reduced system of a Rydberg atom (A) embedded in a bath (B)
consisting of charged particles $c$ , electrons ($c=e$) 
and (singly) charged ions ($c=i$), charge $e_c$, mass $m_c$, particle density $n_c$ and temperature $T$.
The microscopic model under consideration is a hydrogen atom coupled to a surrounding charge-neutral plasma, $\sum_c e_cn_c=0$.
In the bath, in general, the formation of bound states such as atoms is also possible. Furthermore, the 
interaction of the atom is mediated by the Maxwell field which contains, besides the Coulomb interaction 
with the charged particles,
also single-particle states, the photons.
The total system is then described by the Hamiltonian 
\begin{equation}
 \hat H=\hat H_{\rm A}+\hat H_{\rm B}+\hat H_{\rm int}.
\end{equation}
In a plasma environment the Hamiltonian $\hat H_{\rm B}$ includes both the kinetic energy and the 
Coulomb interactions of charged particles $\hat H_{\mathrm{Coul}}$ (see Eq. (\ref{hamiltonian:bath}) below)
as well as the degrees of freedom of the photonic field $\hat H_{\mathrm{photon}}^{\bot}$ 
describing the transversal Maxwell field of the plasma environment,
i.e. $\hat H_{\rm B} = \hat H_{\mathrm{Coul}} + \hat H_{\mathrm{photon}}^{\bot}$. 

The atomic Hamiltonian reads in the non-relativistic case
\begin{equation}
  \label{eq:ham-atom}
   \hat H_{\rm A} = \frac{{\mathbf{\hat P}}^2}{2M}+\frac{{\mathbf{\hat p}}^2}{2m}
  -  \frac{e^2}{4\pi\veo\vert
    {\mathbf{\hat r}}\vert}\,, 
\end{equation}
where the center-of-mass (c.o.m.) motion is described by the total mass $M=m_e+m_i$ and the variables $\bf \hat R,\hat P$, 
the relative motion by the reduced mass $m$ and the relative variables $\bf \hat r,\hat p$. 
The eigenstates $|\Psi_{n,{\bf P}} \rangle$ of the isolated hydrogen atom are the solutions of the Schr\"odinger equation 
$\hat H_{A}|\Psi_{n,{\bf P}} \rangle=E_{n,{\bf P}}|\Psi_{n,{\bf P}} \rangle$ with the eigenenergy 
$E_{n,{\bf P}}= {\mathbf{P}}^2/(2M)+E_{n}$.
The quantum number 
$n=\{\bar n, l,m,m_s\}$ describes the internal state for bound states $E_{n}<0$ and
$n=\{{\bf p},m_s\}$ for scattering states $E_{\bf p}={\bf p}^2/(2m)>0$.
For the bound states, the wave function 
$\Psi({\bf{R,r}})=\langle {\bf{R,r}} |\Psi_{n,{\bf P}} \rangle =\Psi_{{\bf P}}({\bf R}) \psi_n({\bf r})$
contains the eigenstates $\psi_n({\bf r})$ of the hydrogen atom. 
The c.o.m. motion $\Psi_{{\bf P}}({\bf R})$ is given by a plane wave.
In this work we concentrate on the internal degrees of freedom of the bound states.
The c.o.m motion, which, e.g., determines the Doppler broadening of the spectral line profile, will not be discussed 
here in detail. In most cases it will be dropped considering the adiabatic limit.


The interaction between the atomic electron and the plasma environment is given by
the coupling of the atomic current operator to the electromagnetic field of the bath
\begin{equation}
\label{Hint}
  \hat H_{\rm int}(t)=\int d^3 {\bf r} \hat j^{\mu}_{\rm A}(x) \hat A_{\mu,{\rm B}}(x)
\end{equation}
with $x^\mu=\{ct,{\bf r}\}$. 
Introducing the creation (${\hat \psi}^\dagger(x)$) and anihilation (${\hat \psi}(x)$) operator for
the atomic electron, the current operator of the atomic subsystem 
$\hat j^{\mu}_{\rm A}(x)= \{ c  \hat \varrho_{{\rm A}} (x), \hat {\bf j}_{{\rm A}}(x) \}$
can be explicitly written as $ \hat \varrho_{{\rm A}}(x)=-e {\hat \psi}^\dagger(x){\hat \psi}(x)$ 
for the electron probability density and 
$\hat {\bf j}_{{\rm A}} (x)=\frac{i e\hbar}{2 m_e}\left[{\hat \psi}^\dagger(x)
\frac{\partial}{\partial {\bf r}}{\hat \psi}(x) - 
\left(\frac{\partial}{\partial {\bf r}}{\hat \psi}^\dagger(x)\right){\hat \psi}(x) \right]$
for the electric current density of the electron (non-relativistic limit). Without further explanation, the operators
in this work are given in Heisenberg picture
\begin{equation}
 \label{HP}
 \hat O(t)= {\rm e}^{i\hat H t /\hbar} \, \hat O\, {\rm e}^{-i\hat H t /\hbar} .
\end{equation}

The source of the electromagnetic field of the bath 
$\hat A^{\mu}_{{\rm B}}(x)=(\hat U_{{\rm B}}(x),\hat{\mathbf{A}}_{{\rm B}}(x))$ 
is the current density $\hat j^{\mu}_{{\rm B}}(x)$ of all charge carriers in the plasma.
In the present work the Coulomb gauge $\nabla \times \hat{\mathbf{A}}_{{\rm B}}(x))=0$ is used.
The  Fourier transform 
\begin{equation}
\hat{ \mathbf{j}}_{\mathbf{q},{\rm B}}(\omega)
=\int_{\Omega_0}d^3{\bf r}\int_{-\infty}^\infty dt e^{i \omega t-i {\bf q}\cdot{\bf r}}
\hat{ \mathbf{j}}_{{\rm B}}(t,\mathbf{r})
\end{equation}
of the electrical current in the surrounding plasma can be decomposed into a transverse component 
$\sum_c \hat{ \mathbf{j}}^{\bot,c}_{\mathbf{q},{\rm B}}(\omega)$ 
coupled only to the  vector potential $\hat{\mathbf{A}}_{\mathbf{q},{\rm B}}(\omega)$
and a longitudinal one $\sum_c \hat{j}^{||,c}_{\mathbf{q},{\rm B}}(\omega) {\bf q}/q$ which is
related only to the  Coulomb potential. Because of the continuity equation, the relation
${\bf q}\cdot \hat{ \mathbf{j}}_{\mathbf{q},{\rm B}}(\omega) 
=q \hat{j}^{||}_{\mathbf{q},{\rm B}}(\omega) =\omega \hat{\varrho}_{\mathbf{q},{\rm B}}(\omega)$
holds, where $\hat \varrho_{\mathbf{q}}(\omega)$ is the Fourier transform of the corresponding charge density 
operator $\hat \varrho (x)$. 

The general form of the interaction (\ref{Hint}) includes the Coulomb interaction 
via the longitudinal component of the currents, and the coupling of
the transverse component of the currents with the radiation field. 
We do not investigate the radiation
interaction connected with the transverse component.
The radiative field of the plasma determines the natural broadening which 
has already been extensively discussed in{~\cite{BreuerPetruccione,GR13}} by 
using the quantum master equation approach.
However we focus on the Coulomb interaction of the hydrogen atom with its surrounding charged particles in this work.
In this case, the distribution and the motion of the charge carriers in the plasma produce
a scalar potential which is given in terms of the longitudinal current{~\cite{Rein05}}:
\begin{equation} \label{long}
 \hat U_{{\bf q},{\rm B}}(\omega) = \sum_c \frac{\hat \varrho^{c}_{\mathbf{q},{\rm B}}(\omega)}{\epsilon_0 \mathbf{q}^2} 
 = \sum_c  \frac{ \hat j^{||,c}_{\mathbf{q},{\rm B}}(\omega)}{\epsilon_0 \omega q}.
\end{equation}
This results in the pressure broadening of the spectral lines as shown in Sec.{~\ref{sec:lineprofile}}.

The state of the total system is described by the statistical operator $\hat \rho(t)$. 
We assume that the observables $\hat A$ of the 
subsystem A commute with the observables $\hat B$ of the bath B.
If only the properties of the subsystem A are relevant, we can consider the corresponding 
statistical operator 
\begin{equation}
\hat \rho_{\rm A}(t) \equiv {\rm Tr}_{\rm B}\,\hat \rho(t)
\end{equation}
performing the trace over all bath variables.
Then, the average value of any observable $\hat A$ of the subsystem A is calculated as 
$\langle \hat A \rangle^t \equiv {\rm Tr}\{\hat A\ \hat \rho(t)\} ={\rm Tr}_{\rm A}\{\hat A\  \hat \rho_{\rm A}(t)\}$.

The equation of motion for the total statistical operator $\hat \rho(t)$ \cite{GR13} reads
\begin{equation}\label{vonNeumann}
 \frac{\partial }{\partial t} \hat  \rho(t)-\frac{1}{i \hbar}  [ \hat H, \hat \rho(t)]= -\varepsilon [\hat \rho(t)-\hat \rho_{\rm rel}(t)]
\end{equation}
with the relevant statistical operator $\hat \rho_{\rm rel}(t) = \hat \rho_{\rm A}(t) \hat \rho_{\rm B}$
which implies that the quantum systems A and B are uncorrelated.
The equilibrium state $ \hat \rho_{\rm B}$ of the bath B is assumed as the grand canonical distribution 
\begin{equation}
\label{grandcan}
 \hat  \rho_{\rm B}= \frac{1}{Z_{\rm B}}\exp\left[-\frac{ \hat H_{\rm B}-\sum_c\mu_c  \hat N_c}{k_BT}\right], 
 \qquad Z_{\rm B}={\rm Tr}_{\rm B}\exp\left[-\frac{ \hat H_{\rm B}-\sum_c\mu_c  \hat N_c}{k_BT}\right]
\end{equation}
with the 
chemical potentials $ \mu_c $ of the species $c$. 
The limit $\varepsilon \to 0^{+}$ has to be performed after the thermodynamic limit.

A closed equation of motion can be derived for the reduced statistical operator $\hat \rho_{\rm A}(t)$
of the subsystem A by performing the average with respect to the bath in{~\eqref{vonNeumann}}.
If the bath is assumed to have short memory in the sense that the correlation in the bath decays very
quickly in comparison to the time evolution of the reduced system (Markov approximation), and the dynamics
of the reduced system is considered only in second order with respect to $\hat H_{\rm int}$ (Born approximation),
we obtain{~\cite{GR13}} 
\begin{equation}
\label{eq:bamastershmkomS}
 \frac{\partial }{\partial t}  \hat \rho_{\rm A}(t)-\frac{1}{i \hbar}  [ \hat H_{\rm A}, \hat \rho_{\rm A}(t)]= {\cal D}[ \hat \rho_{\rm A}(t)]
\end{equation}
with the influence term
\begin{equation}
\label{influence}
 {\cal D}[ \hat \rho_{\rm A}(t)]=-\frac{1}{\hbar^2}\int_{-\infty}^0 d\tau \,\e^{\varepsilon \tau}\, 
 {\rm Tr}_{\rm B}\left[ \hat H_{\rm int},\left[ \hat H_{\rm int}(\tau), \hat \rho_{\rm A}(t)  \hat \rho_{\rm B} \right]\right].
\end{equation}
This is the quantum master equation (QME) in Born-Markov approximation.
To go beyond the Born approximation, a more general solution has been given in \cite{ZMR97}.

Born approximation indicates that higher orders of the interaction Hamiltonian in the time evolution of the operator{~\eqref{HP}} 
can be dropped. Consequently, the time dependence in Born approximation is given by the interaction picture 
\begin{equation}
 \label{IP}
 \hat O^{\rm I}(t,t_0)= {\rm e}^{i (\hat H_{\rm A} + \hat H_{\rm B} ) (t-t_0) /\hbar} \,  \hat O\,  {\rm e}^{-i (\hat H_{\rm A} + \hat H_{\rm B} ) (t-t_0) /\hbar} .
\end{equation}
At $t=t_0$, the interaction picture coincides with the Schr{\"o}dinger picture. Note that the time of reference $t_0$ is often taken as zero.
In interaction picture, the QME in Born-Markov approximation reads
\begin{equation}
\label{eq:bamastershmkom}
 \frac{\partial }{\partial t}  {\hat \rho}^{\rm I}_{\rm A}(t,t_0)= 
{\cal D}^{\rm I}(t,t_0)\,,
\end{equation}
i.e., only the perturbation determines the time evolution of ${\hat \rho}^{\rm I}_{\rm A}(t,t_0)$ 
(note that $\hat H_{\rm B}$ commutes with $\hat \rho_{\rm A}(t)$).
The influence term in interaction representation follows as
\begin{equation}
\label{influenceI}
 {\cal D}^{\rm I}(t,t_0)=-\frac{1}{\hbar^2}\int_{-\infty}^0 d\tau\, \e^{\varepsilon \tau} \,
 {\rm Tr}_{\rm B}\left[ {\hat H}^{\rm I}_{\rm int}(t,t_0),\left[ {\hat H}^{\rm I}_{\rm int}(t+\tau,t_0), 
{\hat \rho}^{\rm I}_{\rm A}(t,t_0)  \hat \rho_{\rm B} \right]\right].
\end{equation}
In zeroth order with respect to the perturbation, ${\hat \rho}^{\rm I}_{\rm A}(t,t_0)$ is constant, no changing with time $t$.


\subsection{The Influence Term for a Charged Particle System}
\label{sec:geqme:conn}

In this section the master equation for the reduced statistical
operator (\ref{eq:bamastershmkom}) shall be applied to atomic bound
states in a many-particle plasma environment. However, most of the
discussion is valid for a much more general case. 

For the plasma, surrounding the radiating atom, the Hamiltonian is described by 
\begin{equation} \label{hamiltonian:bath} 
 \hat H_{\rm Coul}=\sum_{c, p} \frac{\hbar^2 { p}^2}{2m_c} \hat c_{ p}^\dagger \hat c_{ p}
+\frac{1}{2}\sum_{c,d,{ p}_1{ p}_2,p'_1p'_2}\frac{e_ce_d}{\epsilon_0 \Omega_0 |{\bf p}_1'-{\bf p}_1|^2} 
\delta_{{\bf p}_1+{\bf p}_2,{\bf p}'_1+{\bf p}'_2} \delta_{\sigma_1,\sigma_1'} \delta_{\sigma_2,\sigma_2'} 
\hat c_{{ p}_1}^\dagger \hat d_{{ p}_2}^\dagger \hat d_{{ p}'_2}\hat c_{{p}'_1} \,
\end{equation}
where we used second quantization $\hat c_p, \hat c_p^\dagger$ for free particle states 
$|p \rangle= |{\bf p}, \sigma \rangle$ 
(wave vector and spin) of charge $c$ . The grand canonical equilibrium (\ref{grandcan}) 
contains also the particle number operator
$\hat N_c=\sum_{p} \hat  c_{ p}^\dagger \hat c^{}_{ p}$. The macroscopic state of the bath is 
fixed by the Lagrange multipliers 
$\mu_c$ and $T$. $\Omega_0$ is the volume of the total system. 
Because of charge neutrality $\sum_c e_c \hat N_c\equiv 0$ both $\mu_e, \mu_i$ are related.
The photonic field $\hat H_{\mathrm{photon}}^{\bot}$ is not relevant in our present consideration
which is focussed on the Coulomb interaction with the charged particles of the bath.

The longitudinal part of the interaction Hamiltonian can be extracted from the general form 
$\hat H_{\mathrm{int}}$ (\ref{Hint})
by using the expression{~\eqref{long}} and performing the Fourier transform with respect to 
the time  for the atomic charge 
density operator  
\begin{equation}
{\hat \varrho}_{{\bf q},{\rm A}}^{\rm I}(t,t_0) =\int_{-\infty}^\infty \frac{d \omega}{2 \pi}\, \e^{-i \omega (t-t_0)}\,
 {\hat \varrho}^{\rm I}_{{\bf q},{\rm A}}(\omega)
\end{equation} 
so that
\begin{equation}
 \hat H^{\rm I,\,||}_{\mathrm{int}}(t,t_0)
 =\sum_{{\bf q}} \frac{1}{\epsilon_0 q^2 \Omega_0}  \int \frac{d \omega}{2 \pi} \,\e^{-i \omega (t-t_0)} \,
{ \hat \varrho}^{\rm I}_{{\bf q},{\rm A}}(\omega) {\hat \varrho}^{\rm I}_{-{\bf q},{\rm B}}(t,t_0)
\end{equation}
with $\hat \varrho_{{\bf q},{\rm B}}=\sum_c  \hat \varrho^{c}_{{\bf q},{\rm B}}$ and 
$ \hat \varrho^{c}_{{\bf q},{\rm B}}= \sum_{ p} e_c 
\hat c_{{\bf p}-{\bf q/2},\sigma }^\dagger \hat c^{}_{{\bf p}+{\bf q/2},\sigma}$.
In this work only the contribution of the electrons in the plasma is considered. 
The ionic contribution should be treated in another 
way, see Sec. \ref{sec:Discussion}. Coming back to the influence term (\ref{influenceI}), the factorization of the 
interaction Hamiltonian allows us to perform the average over the bath degrees of freedom separately 
\begin{eqnarray}
\label{influence1}
&&  {\cal D}^{\rm I}(t,t_0)=-\frac{1}{\hbar^2}\int_{-\infty}^0 d\tau \,\e^{\varepsilon \tau} 
 \sum_{q,q'} \frac{1}{\epsilon_0^2 q^2 {q'}^2 \Omega_0^2} \int \frac{d \omega}{2 \pi} \int \frac{d \omega'}{2 \pi}\,
 \e^{ -i (\omega+\omega') (t-t_0) -i \omega' \tau} \nonumber \\ && 
 \times \left\{\left[ {\hat \varrho}^{\rm I}_{{\bf q},{\rm A}}(\omega) {\hat \varrho}^{\rm I}_{{\bf q}',{\rm A}}(\omega')
{\hat \rho}^{\rm I}_{\rm A}(t,t_0) 
 -  {\hat \varrho}^{\rm I}_{{\bf q}',{\rm A}}(\omega') {\hat \rho}^{\rm I}_{\rm A}(t,t_0)  {\hat \varrho}^{\rm I}_{{\bf q},{\rm A}}(\omega)  \right]
 \langle {\hat \varrho}^{\rm I}_{{-\bf q},{\rm B}}(t,t_0) {\hat \varrho}^{\rm I}_{{-\bf q'},{\rm B}}(t+\tau,t_0) \rangle_{\rm B} \right.
 \nonumber \\ && \left.
-\left[ {\hat \varrho}^{\rm I}_{{\bf q},{\rm A}}(\omega){\hat \rho}^{\rm I}_{\rm A}(t,t_0) {\hat \varrho}^{\rm I}_{{\bf q}',{\rm A}}(\omega') 
 - {\hat \rho}^{\rm I}_{\rm A}(t,t_0) {\hat \varrho}^{\rm I}_{{\bf q}',{\rm A}}(\omega') {\hat \varrho}^{\rm I}_{{\bf q},{\rm A}}(\omega)\right]
 \langle {\hat \varrho}^{\rm I}_{-{\bf q'},{\rm B}}(t+\tau,t_0) {\hat \varrho}^{\rm I}_{{-\bf q},{\rm B}}(t,t_0) \rangle_{\rm B} \right\}
\end{eqnarray}
with $\langle\  \cdots\ \rangle_{\rm B} = {\rm Tr}_{\rm B}\left\{ \cdots \hat \rho_{\rm B} \right\}$.
The charge density autocorrelation function 
$\langle {\hat \varrho}^{\rm I}_{-{\bf q},{\rm B}}(t,t_0) {\hat \varrho}^{\rm I}_{-{\bf q'},{\rm B}}(t+\tau,t_0) \rangle_{\rm B}$
is calculated in thermodynamic equilibrium. Because of homogeneity in space and time 
it is $\propto \delta_{{\bf q'},{-\bf q}}$
and not depending on the time $t$ as well as $t_0$.
We introduce the Laplace transform of the bath auto-correlation functions which can be also 
defined as the response function
\begin{eqnarray} \label{Gamma} 
\Gamma_r( \mathbf{q},\omega)  
 = \frac{1}{\hbar^2}\int_{-\infty}^0 d \tau\,  \e^{\varepsilon \tau} \e^{- i \omega \tau} 
 \langle {\hat \varrho}^{\rm I}_{-{\bf q},{\rm B}}(t_0,t_0) {\hat \varrho}^{\rm I}_{{\bf q},{\rm B}}(t_0+\tau,t_0) \rangle_{\rm B}.  
\end{eqnarray}
The response function $\Gamma_r( \mathbf{q},\omega)$ is a complex physical quantity which is 
related to the dynamical structure factor of the plasma or the dielectric function, as shown 
in the App.{~\ref{Kramers-Kronig}}. It can be decomposed into real and imaginary parts,
\begin{equation} \label{realGamma}
  \Gamma_r( \mathbf{q},\omega)   =\frac{1}{2} \gamma_{r}( \mathbf{q},\omega)+i S_{r}( \mathbf{q},\omega),
\end{equation}
where $\gamma_{r}( \mathbf{q},\omega)$ and $S_{r}( \mathbf{q},\omega)$ are both real functions.
They fulfill the Kramers-Kronig relation and are related to the damping and the spectral line shift,
respectively (see Eqs.{~\eqref{imag1}} and{~\eqref{influenceRWA7}} in App.{~\ref{App::RWA}}).

With the response function{~\eqref{Gamma}}, we find that the influence term (\ref{influenceI})
can be rewritten as 
\begin{eqnarray}
\label{influence2}
{\cal D}^{\rm I}(t,t_0)
= -\sum_{q} \frac{1}{\epsilon_0^2 q^4 \Omega_0^2} \int \frac{d \omega}{2 \pi} \int \frac{d \omega'}{2 \pi}
 \e^{ i (\omega'-\omega) (t-t_0) } 
  \Gamma_r(\mathbf{q},- \omega')   \left[{\hat \varrho}^{\rm I}_{{\bf q},{\rm A}}(\omega),{\hat \varrho}^{\rm I}_{-{\bf q},{\rm A}}(-\omega')\hat \rho^{\rm I}_{\rm A}(t,t_0) \right]
 + {\rm h.c.} 
\end{eqnarray} 
The second contribution of the r.h.s. of Eq. (\ref{influence2}) is the hermitean conjugate of
the first contribution so that ${\cal D}^{\rm I}(t,t_0)$ is a real quantity. Approximations for the 
response function $\Gamma_r( \mathbf{q},\omega)$ are obtained from the approximations for the dielectric function such as the random-phase approximation and improvements accounting for collisions.

\subsection{Atomic Quantum Master Equation}
\label{sec:geqme:atom}

In a next step we introduce the orthonormal basis of the hydrogen bound states in the Hilbert
space of the atomic subsystem to obtain the Pauli equation for population numbers and the spectral line profiles.

\subsubsection{Pauli Equation for Occupation Numbers}
\label{sec:geqme:pauli}

We use the basis of hydrogen-like states $|\psi_{n} \rangle$ of the Hamiltonian $\hat H_{\rm A}$. 
For the charge density operator
\begin{equation}
 \hat \varrho_{{\bf q},{\rm A}} =\int d^3 \bar r \,\e^{i {\bf q} \cdot \bar {\bf r}} \hat \varrho_{{\rm A}}(\bar {\bf r}) 
 =\int d^3 \bar r\, \e^{i {\bf q} \cdot \bar {\bf r}}[e_e \delta(\hat {\bf r}_e-\bar {\bf r})+e_i \delta(\hat {\bf r}_i-\bar {\bf r})]
 =e_e\, \e^{i {\bf q} \cdot  \hat {\bf r}_e}+e_i\, \e^{i {\bf q} \cdot  \hat {\bf r}_i} ,
\end{equation}
the time dependence in the interaction picture can be written in matrix representation as ($e_e=-e_i$)
\begin{equation}\label{matrixel}
\hat \varrho_{{\bf q},{\rm A}}^{\rm I}(t,t_0) 
= \e^{\frac{i}{\hbar}\hat H_{\rm A}(t-t_0)}\,\hat \varrho_{{\bf q},{\rm A}}\,\e^{-\frac{i}{\hbar}\hat H_{\rm A}(t-t_0)}
 = \sum_{nn'} e_e \hat T_{n'n} \, F_{n'n}({\bf q})\, {\rm e}^{- i \omega_{nn'}(t-t_0)}
\end{equation}
with 
\begin{align}
 \hat T_{n'n} & = | \psi_{n'} \rangle \langle \psi_{n} | , \\
 \omega_{nn'} & = \frac{ E_{n}-E_{n'} }{\hbar} , \\
 F_{n'n}({\bf q}) & =\int d^3 {\bf r}\, \psi_{n'}^*({\bf r})\psi_{n}({\bf r}) \, (1-e^{-i {\bf q} \cdot {\bf r}}) , 
\end{align}
in adiabatic approximation $m_e \ll m_i$. 
Furthermore, the atom is assumed to be localized at ${\bf R}=0$. 
Performing the Fourier transformation with respect to $t$ we obtain
the atomic charge density in Fourier-space
\begin{eqnarray} \label{rhoomega}
 \hat  \varrho^{\rm I}_{{\bf q},{\rm A}}(\omega)   =  \sum_{nn'} e_e \hat T_{n'n} \, F_{n'n}({\bf q})
\, 2 \pi \, \delta(\omega -\omega_{nn'}) .
\end{eqnarray}

With Eq.{~\eqref{rhoomega}} the influence function{~\eqref{influence2}} can be represented as
\begin{eqnarray} \label{influence7}
{\cal D}^{\rm I}(t,t_0) = - \sum_{nn',mm',{\bf q}} \e^{ - i (\omega_{nn'} + \omega_{mm'} ) (t-t_0) }
 \, K_{mm';n'n}({\bf q},\omega_{mm'})\, \Big\{ \hat T_{n'n} \hat T_{m'm} \hat \rho^{\rm I}_{\rm A}(t,t_0)
   -  \hat T_{m'm}\hat \rho^{\rm I}_{\rm A}(t,t_0) \hat T_{n'n} \Big\} + \rm {h.c.}
\end{eqnarray}
with 
\begin{equation}
K_{mm';n'n}({\bf q},\omega)= \frac{e^2_e}{\epsilon_0^2 q^4 \Omega^2_0}\,
 F^*_{mm'}({\bf q}) F_{n'n}({\bf q}) \, \Gamma_r(\mathbf{q},\omega)
\end{equation}
containing informations about the atomic system, the plasma bath and the interaction between them.
In matrix representation the atomic QME{~\eqref{eq:bamastershmkom}} can be represented as
($|\psi_i \rangle$ - initial state,  $|\psi_f \rangle$ - final state)
\begin{eqnarray}\label{influence10}
 \frac{\partial}{\partial t}\,  \rho_{{\rm A},if}^{\rm I}(t,t_0) = \langle \psi_{i} | {\cal D}^{\rm I}(t,t_0) | \psi_{f} \rangle
\end{eqnarray}
with the influence function
\begin{eqnarray}
 \langle \psi_{i} | {\cal D}^{\rm I}(t,t_0) | \psi_{f} \rangle && 
 = - \sum_{mn,{\bf q}} \Big \{ {\rm e}^{i \omega_{im}(t-t_0)} \, K_{mn;in}({\bf q},\omega_{mn}) \, \rho^{\rm I}_{{\rm A},mf}(t,t_0)
 + {\rm e}^{i \omega_{mf} (t-t_0)} \, K^*_{mn;nf}({\bf q},\omega_{nf}) \, \rho^{\rm I}_{{\rm A},im}(t,t_0) \nonumber \\&& \qquad \qquad
 - {\rm e}^{ i (\omega_{im} + \omega_{nf}  ) (t -t_0)} \, \big[ K_{mi;fn}({\bf q},\omega_{mi}) + K^*_{nf;mi}({\bf q},\omega_{nf}) \big ] 
 \, \rho^{\rm I}_{{\rm A},mn}(t,t_0) \Big \}
\end{eqnarray}
with the density matrix $\rho^{\rm I}_{{\rm A},mn}(t,t_0)=  \langle \psi_{m} |\hat \rho^{\rm I}_{{\rm A}}(t,t_0) | \psi_{n} \rangle$.
The corresponding atomic QME in Schr{\"o}dinger picture is obtained with $\rho_{{\rm A},mn}^{\rm I}(t,t_0) = \e^{i\omega_{mn} (t-t_0)} \rho_{{\rm A},mn}(t)$, 
see Appendix \ref{App::RWA}.

We investigate the diagonal elements of the density matrix 
by setting $i=f$ in the above expression{~\eqref{influence10}}.
This leads to an equation 
for the population number $P_i(t) = \rho^{\rm I}_{{\rm A},ii}(t,t_0)= \rho_{{\rm A},ii}(t)$
\begin{eqnarray}\label{paulibefore}
 \frac{\partial P_i(t)}{\partial t} && = \sum_{n,{\bf q}}
 \Big[ k_{ni}({\bf q},\omega_{ni}) P_{n}(t) - k_{in}({\bf q},\omega_{in}) P_{i}(t) \Big]
   - \sum_{n,m\neq i, {\bf k}} 2 {\rm Re} \Big [ {\rm e}^{  i \omega_{im} ( t-t_0) }\, K_{mn;in}({\bf q},\omega_{mn}) \Big ] 
  \,\rho^{\rm I}_{{\rm A},mi}(t,t_0)  \nonumber \\ && \quad
 + \sum_{m>n,{\bf q}} 2{\rm Re}\bigg\{ {\rm e}^{ i \omega_{nm}  (t-t_0) }\, \Big [ K^*_{ni;mi}({\bf q},\omega_{ni}) 
 + K_{mi;in}({\bf q},\omega_{mi}) \Big ] \bigg\} \,   \rho^{\rm I}_{{\rm A},mn}(t,t_0)
\end{eqnarray}
with $k_{ab}({\bf q},\omega_{ab}) = 2\,{\rm Re} \,\,K_{ab;ab}({\bf q},\omega_{ab})
=e^2_e\, |F_{ab}({\bf q})|^2 \, \gamma_r(\mathbf{q},\omega_{ab})/(\epsilon_0^2 q^4 \Omega^2_0)$, where
expression{~\eqref{realGamma}} is used and the indices $m$ and $n$ are interchanged in the derivation.
The interaction picture shows a slow time dependence in $\rho^{\rm I}_{{\rm A},nm}(t,t_0)$ owing to the influence of the bath,
Eq. (\ref{eq:bamastershmkom}), and a quick time variation due to the factor ${\rm e}^{ i \omega_{nm}  (t-t_0) }$ 
with $\omega_{nm} \neq 0$.
The second and third term oscillate with the characteristic transition frequencies
$\omega_{nm}$ and $\omega_{im}$, respectively. Subsequently their contributions vanish when averaging over relative larger time
interval in comparison with the inverse of the characteristic transition frequencies,
because the population numbers are approximately constant. This is the so-called {\emph{Rotating Wave Approximation}} (RWA).
For the long-term evolution of the reduced system the nondiagonal elements in Eq.{~\eqref{paulibefore}} can be neglected 
and consequently we obtain a
closed rate equation for the population number -- the Pauli equation: 
\begin{equation}\label{PauliEqu}
  \frac{\partial P_i(t)}{\partial t}  = \sum_{n,{\bf q}}
  \Big[ k_{ni}({\bf q},\omega_{ni}) P_{n}(t) - k_{in}({\bf q},\omega_{in}) P_{i}(t) \Big].
\end{equation}
Comparing with the standard form of the Pauli equation
$ \frac{\partial} {\partial t}P_i(t)  = \sum_{n}  [ w_{n \to i} P_n(t)-  w_{i \to n} P_i(t)] $,
we have for the transition rates
\begin{eqnarray}\label{transrate}
 w_{n \to i}=\sum_{\bf q} k_{ni}({\bf q},\omega_{ni}) = \sum_{\bf q} \frac{e^2_e \,|F_{ni}({\bf q})|^2 \, \gamma_r(\mathbf{q},\omega_{ni})}{\epsilon_0^2 q^4 \Omega^2_0}, \quad
 w_{i \to n}=\sum_{\bf q} k_{in}({\bf q},\omega_{in}) = \sum_{\bf q} \frac{e^2_e \, |F_{in}({\bf q})|^2 \, \gamma_r(\mathbf{q},\omega_{in}) }{\epsilon_0^2 q^4 \Omega^2_0}.
\end{eqnarray}
To derive the Pauli equation we used
the RWA which neglects quickly oscillating terms. 
Also the dependence on the time $t_0$ where the interaction picture coincides with the Schr{\"o}dinger picture disappears.
The validity of the RWA in the theory of open quantum systems is under discussion.
The dynamics is modified if contributions of the right hand side of Eq. \eqref{paulibefore} are dropped.
In our investigation we found that if the RWA is carried out prematurely, it will be inappropriate
to describe the dissipative properties of the relevant atomic system (Rydberg states) and result in erroneous transition rates.
More details on that can be found in Appendix{~\ref{App::RWA}.
The nondiagonal elements of Eq.{~\eqref{influence10}} are also discussed there.

\subsubsection{Quantum Kinetic Approach to Spectral Line Profile}
\label{sec:lineprofile}

In open quantum system theory one separates a reduced subsystem out from the total quantum system, which includes all relevant
observables that one is interested in. The remaining degrees of freedom are treated as irrelevant for the
dynamical behavior and are denoted as the observables of the bath. However, 
the selection of the relevant observables that are appropriate to describe
the dynamics of the system depends sensitively on the physical problems that we tackle.

For instance, the degrees of freedom of the emitted photons are irrelevant for the dynamics of the population numbers of the atomic energy
eigenstates and therefore can be considered as part of the bath in the derivation of the Pauli equation. This consideration is also applied
in the derivation of the natural line width of the spectral line profile{~\cite{BreuerPetruccione,GR13}}. In contrast, these degrees of freedom are 
most important for the description of the spectral line profile in a plasma environment where we obtain the spectral
line shapes by measuring the energy of the emitted photons. The emitted photons are therefore relevant degrees of freedom.
To correctly describe the spectral line shapes via the open quantum system theory, we must extend 
the reduced system by including the set of the degrees of freedom of the emitted photons. 
This means that the radiation field together with the atomic system should be considered as the reduced system 
to be described by the QME, and the surrounding plasma is the bath coupled to the system by Coulomb interaction.

Absorption as well as spontaneous and induced emission coefficients, related by the Einstein relation, are obtained from QED 
where the transverse part of the Maxwell field 
\begin{eqnarray}
\hat  H^\perp_{\rm photon}  = \sum_{{\bf k},s} \hbar \omega_{{\bf k},s} \hat n_{{\bf k},s}
\end{eqnarray}
is quantized and denoted by the photon modes $|{\bf k},s\rangle$. 
The frequency $\omega_{\bf k} = c |{\bf k} |= 2 \pi c /\lambda $ is 
the dispersion relation for the frequency as a 
function of the wave number $\lambda$. $\hat n_{{\bf k},s} = \hat b^{\dagger}_{{\bf k},s} \hat b_{{\bf k},s}$ is the occupation number with
the polarization $s=1,2$. As mentioned above, the photon field must be treated as part of the reduced system with the Hamiltonian 
$\hat H_{\rm S} = \hat H_{\rm A} + \hat H^\perp_{\rm photon}$, and the eigenstates will be denoted by the expression 
$|\tilde n\rangle = |\psi_{n},N_{n}({\bf k},s) \rangle$ containing corresponding quantum numbers for the eigenenergy 
$\tilde  E_{n} = E_n + \sum_{{\bf k},s} N_n({\bf k},s)\,\hbar \omega_{{\bf k},s}$ with the occupation number 
$N_n({\bf k},s)$ of the mode $|{\bf k},s\rangle$. 

Emission and absorption are described by the interaction Hamiltonian, see Eq.{~\eqref{Hint}}, 
$\hat H^{\rm rad} = \int d^3 {\bf r}\,{\hat {\bf j}^{\bot}}_{\rm A} \cdot {\hat {\bf A}}_{\rm ph} 
=\int d^3 {\bf r}\, {\hat {\bf d}}_{\rm A} \cdot {\hat {\bf E}}_{\rm ph}$ 
after integration by parts with the atomic dipole operator ${\hat {\bf d}}_{\rm A}$.
The decomposition of the electric field of the photon subsystem (two polarization vectors 
$ \hat {\bf e}_{{\bf k},s}$) is
\begin{eqnarray}
 \hat {\bf E}_{\rm ph} = i \sum_{{\bf k}, s} \sqrt{\frac{\hbar \omega_{\bf k}}{2 \Omega_0}} \, \hat {\bf e}_{{\bf k},s}
 \, [{\hat b}_{{\bf k},s} - {\hat b}^{\dag}\!_{{\bf k},s}] .
\end{eqnarray}
For a given measured photon mode $|\bar {\bf k},\bar s \rangle$ in the experiment, only the mode with ${\bf k}= \bar{\bf k}$
and $s=\bar s$ in the Hamiltonian $\hat H^{\rm rad}$ contributes. This allows us to introduce a new operator
describing emission and absorption
\begin{eqnarray}\label{dipoleOQS}
 \hat {\bf d}_{\rm S} = \hat {\bf d}_{\rm A} \otimes (\hat b_{\bar{ \bf k}} - \hat b^{\dag} \!_{\bar {\bf k}}),
\end{eqnarray}
where the polarization index is suppressed. The initial and final states in this case are given by 
$|\tilde i\rangle = |\psi_{i}, N_{i}({\bf k})\rangle$ and $|\tilde f\rangle= |\psi_{f}, N_{f}({\bf k})\rangle$
with $N_{f}({\bf k}) = N_{i}({\bf k})+\delta_{{\bf k},{\bar {\bf k}}}$, respectively. This means that
for the measured photon mode ${\bar {\bf k}}$ the occupation number fulfills  
$N_{f}({\bar {\bf k}}) = N_{i}({\bar {\bf k}})+1$, 
while for all other photon modes their occupation numbers remain unchanged.
A shift of the eigenenergy levels is caused by the interaction with the plasma environment via the momentum exchange.
Subsequently, this leads to a deviation of the measured transition frequency $\omega_{\bar{ \bf k}}$ from the 
characteristic transition frequencies $\omega_{nn'}$ between the unperturbed atomic eigenstates $|\psi_n \rangle$.
We define the deviation by using the eigenenergies $\tilde  E_{n}$ via
\begin{equation}
 \Delta \omega_{nn'} = (\tilde E_{n}- \tilde E_{n'})/\hbar.
\end{equation}

We use the interaction picture with $\hat  H_{0} = \hat H_{\rm S} +\hat H_{\rm B}$ so that the power spectrum 
$P(\omega_{\bar{ \bf k}}) = \int_0^\infty  e^{-\epsilon t} e^{ \pm i \omega_{\bar{ \bf k}} t} \langle \hat {\bf d}_{\rm A} \rangle^t dt$
as shown in{~\cite{BHMV84}} in the framework of the linear-response theory can be rewritten as
\begin{equation} \label{powernew}
P(\omega_{\bar{ \bf k}}) = \int_0^\infty  e^{-\epsilon t} \langle {\hat {\bf d}_{\rm S}} \rangle^t dt 
= \sum_{if} {\cal L}_{i,f},
\end{equation}
where the photon frequency is absorbed by the new dipole operator $\hat {\bf d}_{\rm S}$ of the reduced system (including photons) and 
\begin{equation} \label{dipoletime}
\langle {\hat {\bf d}_{\rm S}} \rangle^t ={\rm Tr} \{\hat \rho_{\rm S}(t)  \hat {\bf d}_{\rm S} \} 
 =\sum_{if} \rho^{\rm I}_{{\rm S},fi}(t) {\bf d}^{\rm I}_{{\rm S},if}(t)
\end{equation}
with $\rho^{\rm I}_{{\rm S},fi}(t)$ being the solution of the QME in interaction picture (see Eq.{~\eqref{profileQME}}),
and the matrix elements 
${\bf d}^{\rm I}_{{\rm S},if}(t) =\langle \psi_i|{\bf d}_{\rm A}|\psi_f \rangle e^{- i \Delta \omega_{if} t}$.
Consequently, the spectral line shape ${\cal L}_{i,f}$ in Eq.{~\eqref{profileQME}} can be written as
\begin{equation} \label{shapefunc}
 {\cal L}_{i,f} = \int_0^\infty dt\,  e^{-\epsilon t}\, \rho^{\rm I}_{{\rm S},fi}(t) {\bf d}^{\rm I}_{{\rm S},if}(t).
\end{equation}

In order to obtain the solution of the QME, a similar reduced charge density operator containing the photon information as in Eq.{~\eqref{dipoleOQS}}
can be introduced for the extended reduced system
\begin{eqnarray}
\hat \varrho_{{\bf q}, {\rm S}} = \hat \varrho_{{\bf q}, {\rm A}} \otimes (\hat b_{\bar{ \bf k}} - \hat b^{\dag}_{\bar {\bf k}}).
\end{eqnarray}
Using the basis set $|\tilde n \rangle$ of the unperturbed reduced system, we obtain the matrix
elements of the reduced charge density operator $\hat \varrho^{\rm I}_{{\bf q}, {\rm S}} (t)$ at time $t$
\begin{eqnarray}
 \langle \tilde n' |\hat \varrho^{\rm I}_{{\bf q}, {\rm S}} (t) |\tilde n \rangle
 = e_e \, F_{n'n}({\bf q})\,  e^{i \Delta \omega_{n'n} t } \, 
 \left[ \delta_{N_{n'}({\bar{ \bf k}}), N_{n}({\bar{ \bf k}})-1} 
 - \delta_{N_{n'}({\bar{ \bf k}}), N_{n}({\bar{ \bf k}})+1} \right], \nonumber
\end{eqnarray}
where the Kronecker's delta is connected to the atomic emission and absorption with the 
transition frequency $\omega_{n'n}$. 
Performing the Fourier transform with respect to the time $t$, we obtain
the reduced charge density operator in Fourier-space 
\begin{eqnarray}
  \hat \varrho_{{\bf q}, {\rm S}} (\omega) 
 = \sum_{n'>n} e_e\, \hat T_{n'n}^{-}  \cdot F_{n'n} ({\bf q})\,  \delta(\omega - \Delta \omega_{n'n})
 - \sum_{n'<n} e_e\,\hat T_{n'n}^{+} \cdot F_{n'n} ({\bf q}) \, \delta(\omega +\Delta \omega_{nn'}) 
\end{eqnarray}
with $\hat T_{n'n}^{-} = |\tilde n' \rangle \langle \tilde n | \cdot \delta_{N_{n'}({\bar{ \bf k}}), N_{n}({\bar{ \bf k}})-1}$ 
denoting the one photon absorption
and $ \hat T_{n'n}^{+} = |\tilde n' \rangle \langle \tilde n | \cdot \delta_{N_{n'}({\bar{ \bf k}}), N_{n}({\bar{ \bf k}})+1}$ 
- the one photon emission.

The QME in RWA in interaction picture can be written in terms of the matrix element
$\rho_{{\rm S},fi}^{\rm I}(t) =\langle \tilde f |\hat \rho_{{\rm S}}^{\rm I}(t)|\tilde i \rangle $:
\begin{eqnarray}\label{profileQME}
 \frac{\partial \rho_{{\rm S},fi}^{\rm I}(t)}{\partial t} = -  \Gamma_{fi}^{{\rm BS}}(\omega_{\bar {\bf k}}) \rho_{{\rm S},fi}^{\rm I}(t)
 + \Gamma_{fi}^{\rm V}\,\rho_{{\rm S},fi}^{\rm I}(t),
\end{eqnarray}
which is shown in detail in Appendix{~\ref{App::Profile}}.
The influence function, the right side of the Eq.{~\eqref{profileQME}}, characterizes the spectral intensity of the emitted 
photons by a coefficient $\Gamma_{fi}^{BS}(\omega_{\bar {\bf k}}) $ describing the shift of the eigenenergy levels
and the pressure broadening
\begin{eqnarray}
 \Gamma_{fi}^{{\rm BS}}(\omega_{\bar {\bf k}})
 = \sum_{n,{\bf q}} \big\{ K_{nf;fn}(\mathbf{q},  \Delta \omega_{nf}) + K_{nf;fn}(\mathbf{q}, - \Delta \omega_{fn})
 + K^{*}_{ni;in}(\mathbf{q},  \Delta \omega_{ni}) +  K^{*}_{ni;in}(\mathbf{q}, - \Delta \omega_{in})\big \}
\end{eqnarray}
and a coefficient $\Gamma_{fi}^{\rm V} $ describing the vertex correction
\begin{eqnarray}
\label{eq:vertex}
  \Gamma_{fi}^{\rm V}
  = \sum_{{\bf q}} \big\{ K_{ii;ff}(\mathbf{q},  \Delta \omega_{ff}) + K_{ii;ff}(\mathbf{q}, - \Delta \omega_{ff}) 
  + K^{*}_{ff;ii}(\mathbf{q},  \Delta \omega_{ii}) +  K^{*}_{ff;ii}(\mathbf{q}, - \Delta \omega_{ii}) \big\} ,
\end{eqnarray}
The vertex correction has no dependence on the photon frequency $\omega_{\bar {\bf k}}$ and contributes only beyond
the dipol approximation.
Formally integrating the expression {\eqref{profileQME}} yields
\begin{eqnarray}
 \rho_{{\rm S},fi}^{\rm I}(t) = \rho_{{\rm S},fi}^{\rm I}(0) 
 \cdot e^{- \left\{ \Gamma_{fi}^{{\rm BS}}(\omega_{\bar {\bf k}}) - \Gamma_{fi}^{\rm V} \right\} t}.
\end{eqnarray}
Inserting this formal solution into the Eq.{~\eqref{shapefunc}}, the line shape function can be expressed as
\begin{eqnarray}
\label{shape}
 {\cal{L}}(\omega_{\bar{ \bf k}})_{i,f} && \propto 
 \frac{1 }{  \omega_{\bar{ \bf k}} -  \omega_{if}  + i \epsilon 
 - i \Gamma_{if}^{{\rm BS}}(\omega_{\bar {\bf k}})  + i \Gamma_{if}^{\rm V} }.
\end{eqnarray}
The expression{~\eqref{shape}} coincides with the result of the unified theory for spectral line 
profiles \cite{Guen95} if only the electron contribution (impact approximation) is considered.
Note that the unified theory gives the result in Born approximation with respect to the 
interaction with the surrounding plasma, what corresponds to the Born-Markov approximation for the 
coupling to the plasma
considered as the bath. Strong coupling of the radiator to the perturbing environment has been treated in the 
Green function approach using a T-matrix approximation, see \cite{Guen95}. The improvement of the 
Born-Markov approximation for the 
QME considering strong interactions and the ionic contribution of the plasma environment,
given by the microfield distribution, will be discussed below in Sec. \ref{sec:Discussion}.

\section{Robust circular Wave Packet and Transition Rates}
\label{sec:wavepa}
An advantage of the QME is the possibility to introduce optimal (robust) states which allows
the transition from a quantum description to a classical one. In the case of Rydberg atoms,
one considers electrons in highly excited hydrogen states. With increasing quantum number $\bar n$, 
a pure hydrogen orbital can be formed only if the atom is well isolated from external influences.
When the interaction with the bath is comparable or greater than the 
differences of atomic energy eigenstates $E_{n}$ for $\bar n$ near a fixed value $n_0$,
the wave packet 
description is more appropriate to describe the evolution of the system,
in particular transition rates. For a local interaction such as the Coulomb potential, 
the position $\bf r$ of the atomic electron enters the interaction part of the Hamiltonian, 
and localization is favored because $\bf r$ commutes with $\hat H_{\rm int}$ and is a conserved 
quantity with respect to this part of the Hamiltonian.

In addition, the introduction of the wave packet description may allow us 
to investigate the boundary between the quantum and classical 
descriptions of systems. In fact,
since the introduction of quantum mechanics many physicists attempted 
to establish the connections between these
descriptions of nature by exhibiting the so-called coherent
wave packet. One of the famous examples is the well known
coherent state of the linear harmonic oscillator{~\cite{Glauber63}} 
which may be regarded 
as an excellent example to describe the macroscopic limit of a quantum 
mechanical system according to the correspondence principle. For 
the Coulomb problem, e.g. the hydrogen atom, many attempts to construct
localized semi-classical solutions of the coherent-state type
have been made{~\cite{Brown73,NS78,BRG86,ZZF94,MS97}}.
Note that the hydrogen atom is equivalent to the four-dimensional harmonic
oscillator so that coherent wave packets can be introduced accordingly{~\cite{BRG86}}.
Recently, Makowski and Peplowski constructed very well-localized two-dimensional wave packets  
for two different potentials{~\cite{MP12,MP13}} where a very good quantum-classical 
correspondence is observed. In the present paper we use Brown's circular-orbit
wave packet{~\cite{Brown73,GS90}} as a quasiclassical representation to describe the highly
excited Rydberg states of the hydrogen atom.

\subsection{Wave Packet for Circular Motion}
\label{wp:robust}

Within the QME approach, the state of the relevant system $\hat \rho_{\rm S}$ is of
interest, and we can represent the statistical
operator by the density matrix $\rho_{{\rm S},mn}= \langle m|\hat \rho_{\rm S}|n \rangle$
with respect to the states $|n \rangle$ of the
system. For the representation, one can use the orthonormal
basis set of energy eigenstates of the unperturbed bound
system according to the interaction picture. In the case considered here
these are the hydrogen orbitals. For a complete orthonormal
basis the scattering states must be also included.
The hydrogen orbitals are long-living
if the perturbation by the surrounding plasma is small. If the
broadening of the energy levels remains small compared to the
distance between neighbored energy eigenvalues, the transition rate due
to collisions with the plasma is small.

At high excitation (Rydberg states), the interaction effects are no
longer small compared to the level distance, and the
pure quantum state has only a short life time.
Therefore one can look
for more robust states that are formed as superposition
of energy eigenstates but are more stable in the time evolution. In
particular, the Coulomb interaction contains the position operator,
and localized states are more robust with respect to the interaction
with the surrounding plasma. To find the robust states one has to
optimize the quantum states.
For simplicity, we restrict ourselves to circular-orbit eigenfunctions of the hydrogen atom.
In this section, we use the notation $n$ for the principal quantum number,
\begin{eqnarray} \label{sphcoor}
 \psi_{n}({\bf r}) = \langle \mathbf{r} | \psi_{n,n-1,n-1} \rangle = c_n \left(\frac{r}{a_{\rm B}}\right)^{n-1}
 \mathrm{e}^{-r/(na_{\rm B})} \sin^{n-1}(\theta) \mathrm{e}^{i(n-1) \phi},
\end{eqnarray}
where $c_n = (2/(na_{\rm B}))^{3/2}[2n(2n+1)!]^{-1/2} $ denotes the normalization constant.
Furthermore, in this section we use the abbreviation $\psi_{n}({\bf r})$ for the circular wave 
function $\psi_{n,n-1.n-1}({\bf r})$.
It could be seen from Eq.{~\eqref{sphcoor}} that the hydrogen electron in this eigenstate 
is already excellently localized in the radial ($r$) and polar ($\theta$)-direction.
To achieve the localization with respect to the azimuthal direction angle $\phi$, the wave packet can
be introduced.

The circular-orbit wave packet of the hydrogen atom is a coherent state
constructed from the superposition of circular-orbit eigenfunctions
of the hydrogen atom with a Gaussian weighting function around a large 
principal quantum number $n_0${~\cite{GS90}}:
\begin{equation} \label{wpkt}
 |G_{n_0,\phi_{0}} \rangle=  \sum_n \frac{g_{n_0,n}}{\sqrt{{\cal N}_{n_0}}}  e^{i (n-1) \phi_{0} } |\psi_{n} \rangle
\end{equation}
with the Gaussian factor and the normalization factor respectively
\begin{equation} \label{gaussiancoeff}
g_{n_0,n} =  \exp\left\{- \frac{(n-n_0)^2}{4\sigma_{n_0}^2} \right\},\ 
{\cal N}_{n_0} = \sum_{n=1}^{\infty} \exp\left\{-\frac{(n-n_0)^2}{2\sigma_{n_0}^2}\right\} ,
\end{equation}
where $\sigma_{n_0}$ is the standard deviation considered as fixed parameter 
for $n_0$. 
Without loss of generality we can put $\phi_{0} = 0$ because it fixes, as a phase factor, only
the initial position of the wave packet at the azimuthal angle $\phi$. We drop $\phi_{0}$ in the following.
Due to the superposition with a Gaussian factor, we have also good localization with 
respect to $ \phi$
in the wave packet description{~\eqref{wpkt}}.
The actual Hilbert space ${\cal H}_{n,n-1.n-1}$ considered here is only a 
subspace of the entire Hilbert space ${\cal H}$ of the hydrogen atom. 
The generalization to the full Hilbert space to include all bound and scattering states
could be done straightforwardly.

The time-dependent wave packet in the coordinate representation in terms of 
spherical coordinates is given by
\begin{eqnarray}{\label{timeWP}}
 \langle {\bf r} | G_{n_0} \rangle^t =  \sum_n \frac{g_{n_0,n}}{\sqrt{{\cal N}_{n_0}}}\, {\rm e}^{i E_n t/\hbar }\, 
 \psi_{n}({\bf r})
\end{eqnarray}
with $E_n= {\rm Ry}/ n^2$ and ${\rm Ry} = 13.6\, {\rm eV}$.
For an appropriate Gaussian factor only the terms with principal quantum number adjacent to $n_0$ contribute.
Therefore we can use the central quantum number $n_0$ to approximate other states in radial and $\theta$-direction.
In addition, for short-term time evolution the energy $E_n$ in the factor ${\rm e}^{i E_n t/\hbar}$ 
in Eq.{~\eqref{timeWP}} can be expanded around $n_0$
up to the second order, which relates directly to the quantum revival, see below.
The probability distribution of the wave packet can be represented as
\begin{eqnarray}
 |\langle {\bf r} | G_{n_0} \rangle^t|^2 = \sum_{m,n} c^2_{n_0} \left(\frac{r}{a_{\rm B}}\right)^{2{n_0}-2}
 \mathrm{e}^{-2r/(n_0 a_{\rm B})} \sin^{2{n_0}-2}(\theta) 
 \cdot \mathrm{e}^{-(a_1 - i \omega_{\rm rev}t)(n-n_0)^2 -(a_1 + i \omega_{\rm rev}t)(m-n_0)^2 
 + i (\phi - \omega_{\rm cl}  t) (n-m)}
\end{eqnarray}
with $a_1 = 1/(4 \sigma^2_{n_0}) ,\, \omega_{\rm cl} = |E^{'}_{n_0}|/\hbar=  2{\rm Ry}/ (\hbar n_0^3)$ and
$ \omega_{\rm rev} = |E^{''}_{n_0}|/(2 \hbar)= 3{\rm Ry}/ (\hbar n_0^4)$,
where $E^{'}_{n_0}$ and $E^{''}_{n_0}$ are the first and second derivatives of $E_n$ with respect to 
the main quantum number $n$ at $n_0$, respectively. 
As pointed out in{~\cite{Rob04}}, $\omega_{\rm cl} $ relates to the classical Kepler period 
$T_{\rm cl} = 2 \pi r_{\rm cl}/v_{\rm cl}$ for the Kepler trajectory with $r_{\rm cl} = n^2_0\, a_{\rm B}$
and $v_{\rm cl} = \sqrt{{\rm Ry}/(m_e n^2_0)}$,
and the quantum revival period can be defined by $\omega_{\rm rev}$. 

For highly excited states $|x| \ll n_0$ with $x=n - n_0$ and $|y| \ll n_0$ with $y = m - n_0$, the sum $\sum_{m,n}$ 
can be replaced by the integral $\int_{-\infty}^{\infty} d x \int_{-\infty}^{\infty} d y$.
Integrating over the variables $r$ and $\theta$ and performing the integral over $x,\, y$ 
yields the probability distribution of the wave packet
\begin{eqnarray}\label{ProbDist}
 | G_{n_0}(\phi,t)|^2  \sim \sqrt{\frac{\pi^2}{a_1^2 + (\omega_{\rm rev} t)^2 }} 
 \exp\bigg[- \frac{\phi_{\rm cl}^2(t) }{2 [a_1^2 + (\omega_{\rm rev} t)^2]/a_1} \bigg]
\end{eqnarray}
with $ \phi_{\rm cl}(t) = \phi - \omega_{\rm cl}  t$. 
From this probability distribution the time-dependent width of the wave packet for a Rydberg electron can be extracted 
\begin{eqnarray}\label{phiwidth}
 \sigma_{n_0}^{\phi}(t) = \sqrt{  [a_1^2 + (\omega_{\rm rev} t)^2]/a_1} = \sqrt{\frac{1}{4 \sigma^2_{n_0}} +
 \frac{ \sigma^2_{n_0} (E^{''}_{n_0} t)^2}{\hbar^2}}.
\end{eqnarray}
For the initial time $t=0$, we have $\sigma^{\phi}_{n_0} = 1/(2\sigma_{n_0})$.
The expression{~\eqref{ProbDist}} also shows that on such a short time scale the central position 
of the probability distribution is exactly determined by the Kepler motion.
The localized wave packet for the hydrogen atom moves along the 
classical Keplerian trajectory of the electron and its width broadens.
With time evolution the localization of the wave packet is destroyed
and interference fringes of different eigenstates are displayed. 
On a much longer time scale $T_{\rm rev} = 2\pi /\omega_{\rm rev}$, the wave packet finally 
reverses itself, which is the above mentioned quantum revival as indicated in Eq.{~\eqref{ProbDist}}.

The dynamics of the wave packet shown above is purely due to quantum mechanical evolution without 
plasma surroundings. Within a plasma environment, the hydrogen atom undergoes interactions
with the plasma particles which results in the shift of the eigenenergy levels, the broadening
of plasma spectral lines, the screening of the Coulomb potential, the localization of the hydrogen
atom (proton and bound electron), etc. In this work we concentrate on the localization of the
bound electron immersed in a plasma environment.

The scattering of the bound electron by free plasma electrons results in
the localization of the electron of the hydrogen atom, 
i.e. the collisions with the plasma tend to localize the Rydberg 
electron and to narrow the wave packet. As in the case of free
particles in a surrounding environment{~\cite{JZKGKS03}}, the spreading
of the wave packet competes with the localization effect induced by the plasma 
environment. The optimum width of a Gaussian wave packet where both
effects, localization and quantum diffluence, nearly compensate, describes a state
which is nearly stable in time and is denoted as robust state.

In this work we are interested in time scales, which are even
smaller than the classical Keplerian periodicity $ T_{\rm cl}$. 
We assume that on such a short time scale a Rydberg electron behaves like a free electron because of
the weak coupling between the electron and the proton.
Comparing with the relaxation processes, which describe the inelastic 
coupling between the internal energy eigenstates
and the surrounding environment, the quasiclassical Kepler motion 
of the wave packet is assumed to be influenced by the elastic
scattering of the Rydberg electrons with its surroundings.
Similarly as in the case of the quantum Brownian motion of a free particle{~\cite{JZKGKS03}},
the equation of motion for the reduced density matrix with respect to variable 
$x = r_{\rm cl}\, \phi$ obeys
\begin{eqnarray}\label{phimotion}
 \frac{\partial}{\partial t}  \rho(x, x',t) 
 = - \Lambda_{R_n}  (x - x')^2
\end{eqnarray}
with the localization rate 
\begin{equation}\label{localrate}
 \Lambda_{R_n} = \frac{n_{\rm pl}}{\pi (2 \pi \hbar)^2 }  \int_{0}^{\infty} d q\, \frac{q^4}{3}  V^2_{\bf q} \, F^2_{nn}({\bf q})
 \,\sqrt{\frac{m_e}{2 \pi k_B T q^2} }\,{\rm e}^{-\hbar^2 q^2/(8 m_e k_B T)},
\end{equation}
describing how fast 
interferences of an entangled system of extension $|x - x'|$ are suppressed, 
for details see Appendix{~\ref{colldecoh}}.
According to{~\cite{JZKGKS03}}, the optimal width of the wave packet is defined by equilibrating 
the interplay between the spreading of the wave packet and the localization of the wave packet and reads:
\begin{equation}\label{opwidth}
 \sigma^{\rm cl}_{n}=\frac{1}{2} \left( \frac{\hbar}{m_e \Lambda_{R_n}} \right)^{1/4} .
\end{equation} 
As a consequence, an optimal width $\sigma_{n_0}$ can be calculated using the relation{~\eqref{phiwidth}} for $t=0$ 
and the relation $\sigma^{\phi}_{n_0} = \sigma^{\rm cl}_{n_0} / r_{\rm cl}$ so that
\begin{equation}\label{opwavepacket}
 \sigma_{n_0}  = \frac{ r_{\rm cl} }{ 2\, \sigma^{\rm cl}_{n_0} }  .
\end{equation}
For the plasma with temperature $T=300$K and density $n_{\rm pl}=10^{9}{\rm cm^{-3}}$, we obtain an optimal width
$\sigma_{n_0}=0.75$ for $n_0 = 13$, which will be shown in the next section 
to be appropriate to describe the transition rate.

In table \ref{decohrateDensTemp} we show the dependence of the localization rates on the plasma parameters
for different principal quantum numbers. For given temperature and density, the localization rate
decreases slightly with the increasing quantum number $n_0$. At a fixed temperature, the localization rate
is raised drastically when the plasma density increases. At the same time, the localization rate shows only a weak
dependence on the plasma temperature.


\begin{table}[ht]
 \begin{tabular}{|c||c||c|c|c|c|}
 \hline
 {\multirow{2}{*}{T [K]} } &{\multirow{2}{*}{$n_{\rm pl} [{\rm cm^{-3}}]$}}  & \multicolumn{4}{ |c| }{quantum number $n_0$} \\ \cline{3-6}
& & \phantom{aa} 10 \phantom{aa} & \phantom{aa} 20 \phantom{aa} & \phantom{aa} 30 \phantom{aa} & \phantom{aa} 40 \phantom{aa} \\ \cline{1-6}
{\multirow{3}{*}{100} }& $10^{9}$ & $5.722$ & $5.722 $ & $5.722$ &  $5.721$  \\
 & $10^{12}$ & $180.3 $ & $180.1$ & $180.0$ & $179.7$ \\
 & $10^{15}$ & $5090$ & $4969$ & $4813$ & $4645$ \\ \cline{1-6}
{\multirow{3}{*}{1000} }& $10^{9}$ & $5.722$ & $5.722$ & $5.722$ & $5.722$ \\ 
 & $10^{12}$ & $180.9$ & $180.8$ & $180.7$ & $180.6$ \\
 & $10^{15}$ & $5619$ & $5549$ & $5474$ & $5399$ \\ \cline{1-6}
{\multirow{3}{*}{10000} }& $10^{9}$ & $5.722$ & $5.722$ & $5.722$ & $5.722$ \\ 
 & $10^{12}$ & $180.9$ & $180.9$ & $180.9 $ & $180.9$ \\
 & $10^{15}$ & $5696$ & $5670$ & $5645$ & $5619$ \\ \cline{1-6}
\end{tabular}
\caption{Localization rate $\Lambda_{R_n}${~\eqref{localrate}} in units $[10^{23}{\rm cm^{-2} s^{-1}}]$ for different plasma densities and temperatures.}
\label{decohrateDensTemp}
\end{table}

The transition between descriptions of the bound electron in hydrogen atom by the wave packet and the pure quantum 
eigenstate may be determined by comparing the optimal width{~\eqref{opwidth}} with 
the orbit radius $r_{n_0}=r_{\rm cl} = n_0^2 a_{\rm B}$. For this, a function $b(n_0,n_{\rm pl},T)$ 
\begin{equation}
b(n_0) = b(n_0,n_{\rm pl},T) = \frac{ \sigma^{\rm cl}_{n_0}}{r_{n_0}}-1
\end{equation}
can be introduced.
For the given electron density $n_{\rm pl}$ and temperature $T$ of the plasma, quantum mechanical
descriptions is valid for $b(n_0)>0$,
and for the opposite case ($b(n_0)<0$), the wave packet descriptions can be used.

\begin{figure}[ht]
 \centering
\includegraphics[width=0.45\textwidth]{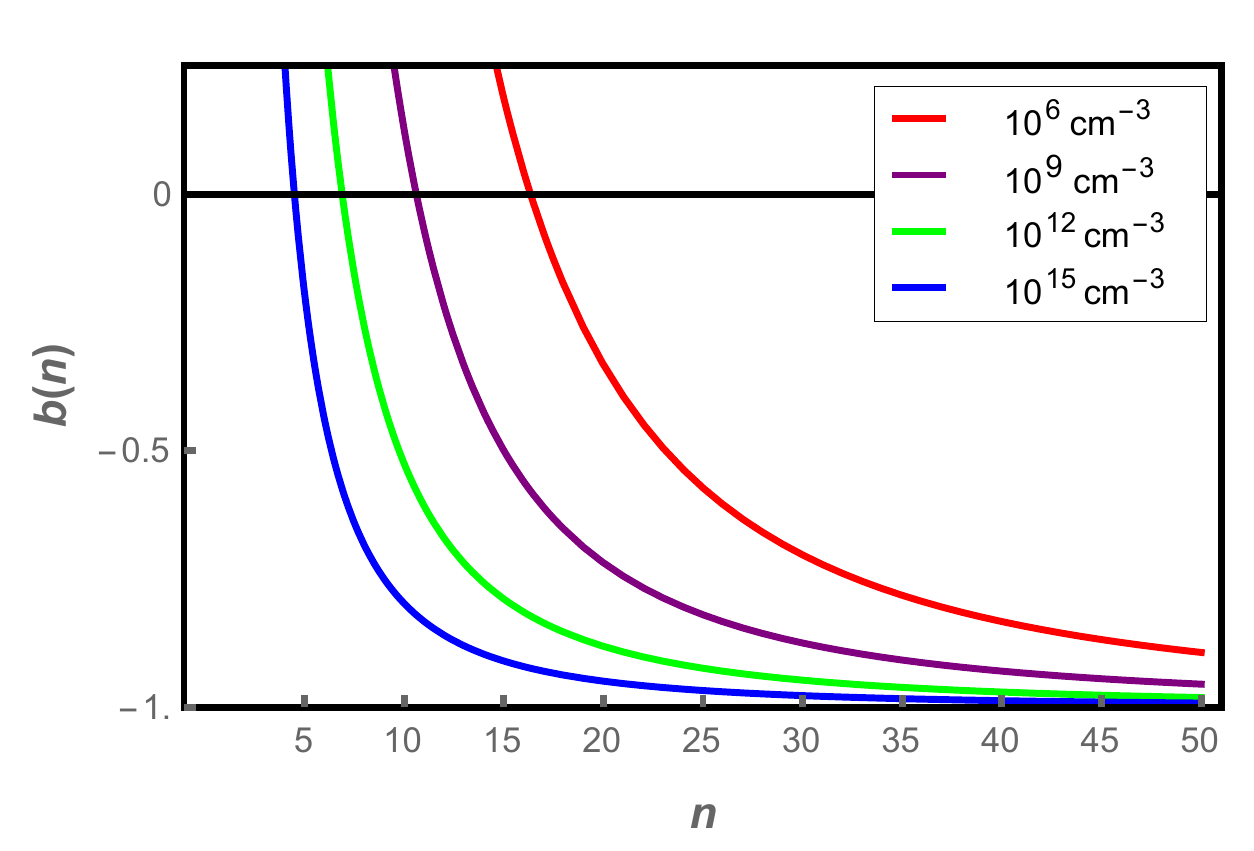}
 \caption{Boundary between classical and quantum mechanical descriptions of the hydrogen electron at $T=300$ K
 for different densities.}
 \label{boundary}
\end{figure}
In Fig.{~\ref{boundary}} we show this function for different plasma densities at the given
temperature $T=300$ K. With the increase of the plasma density the principal quantum number 
$n_{\rm cr}$ at $b(n_{\rm cr})\approx0$ 
characterizing the change from
a pure quantum description to a classical description drops drastically.

We discussed the descriptions of the bound electron, in particular the validity of the wave packet description.
We have shown that this question is  related to the localization
of the wave packet if an optimal width of the wave packet is assumed,
which also has a dependence on the mass
of the localized object as shown in Eq.{~\eqref{opwidth}}. 
Similar considerations can be made for the free electrons and ions in the
plasma (see. Sec.{~\ref{sec:Discussion}}).

\subsection{Transition Rates}
\label{wp:transrate}
We are interested in a matrix representation of the QME.
We use robust states $| i \rangle = |G_{n_i} \rangle$ for the initial state and $| f \rangle = |G_{n_f} \rangle$
for the final state to investigate the atomic transition rates of the Rydberg states. 
For the reduced Hilbert space ${\cal H}_{n,n-1.n-1}$ used to construct the circular-orbit wave packet, 
there is no completeness relation $\sum_n |n\rangle \langle n|= \hat 1$ because non-circular orbits are missing. 
Only if we project on the reduced Hilbert space, this relation can be applied.
A more general discussion about the completeness relation in the wave packet case is 
found in Refs.{~\cite{Fox99,Kl96}}. Therewith the charge density operator in Hilbert space 
${\cal H}_{n,n-1.n-1}$ is expressed as
\begin{eqnarray}
 \hat \varrho_{{\bf q},{\rm A}}^{\rm I} (t)  = \sum_{mn} e_e\, \hat T_{mn}\, F_{mn}({\bf q}) e^{i\omega_{mn} t}.
\end{eqnarray}
In Fourier-space the charge density operator reads
\begin{eqnarray}\label{chargeWP}
\hat \varrho_{{\bf q},{\rm A}}^{\rm I} (\omega)  = \sum_{mn} e_e\,  \hat T_{mn}\, F_{mn}({\bf q})\, 
2 \pi \delta(\omega + \omega_{mn}).
\end{eqnarray}
Note that the operators given in this section are all projected on the reduced Hilbert space ${\cal H}_{n,n-1.n-1}$.
The use of the full Hilbert space is more complex and should be worked out in future investigations.

In the present section, the diffusion of the wave packet with the center quantum number $n_0$ is of essential interest.
The dynamics along the classical trajectory, shown in the previous section, is given by $\phi_{\rm cl}(t)$. 
To investigate the diffusion of the wave packet with respect to the quantum number $n$, we come back to the QME 
in which the influence function for the wave packet in Hilbert space ${\cal H}_{n,n-1,n-1}$ is
obtained by inserting the charge density operator {\eqref{chargeWP}} into equation {\eqref{influence2}}
\begin{eqnarray}
 {\cal D}^{\rm I}[\hat \rho^{\rm I}_{\rm A}(t)]
  =  - \sum_{nn',mm',{\bf q}} \e^{ - i (\omega_{nn'} + \omega_{mm'} ) (t-t_0) }
 \, K_{mm';n'n}({\bf q},\omega_{mm'})\, \Big\{ \hat T_{n'n} \hat T_{m'm} \hat \rho^{\rm I}_{\rm A}(t,t_0)
   -  \hat T_{m'm}\hat \rho^{\rm I}_{\rm A}(t,t_0) \hat T_{n'n} \Big\} + \rm {h.c.} 
\end{eqnarray}
The influence function in RWA can be represented in matrix representation as
\begin{eqnarray}\label{matrixelementWP}
 \langle f | {\cal D}^{\rm I}[\hat \rho^{\rm I}_{\rm A}(t)] | i \rangle && = \sum_{s,h,{\bf q}} 
 {\cal G}^{n_i,h}_{n_f,s} \,  \Big\{ K_{ss;hh}({\bf q}, \omega_{ss}) + K^*_{hh;ss}({\bf q}, \omega_{hh}) \nonumber \\ &&
 - \sum_{n} \big\{ K_{sn;sn}({\bf q}, \omega_{sn}) + K^*_{hn;nh}({\bf q}, \omega_{hn}) \big\}  \Big\} 
 \rho^{\rm I}_{{\rm A},sh}(t)
\end{eqnarray}
with 
\begin{equation}
 {\cal G}^{n_i,h}_{n_f,s}= \frac{g_{n_f,s} \cdot g_{n_i,h}}{\sqrt{{\cal N}_{n_f} {\cal N}_{n_i}}},
\end{equation}
where the $g$-function is given by Eq.{~\eqref{gaussiancoeff}}.
After decomposition of the response function 
$\Gamma_r(\mathbf{q},\omega) = \gamma_r(\mathbf{q},\omega)/2 + i S_r(\mathbf{q},\omega)$,
we have the dissipator for the circular wave packet
\begin{eqnarray} \label{transitionWP}
 \langle f | {\rm D}^{\rm I}[\hat \rho^{\rm I}_{\rm A}(t)] | i \rangle 
 = - \frac{1}{2} \sum_{s,h} {\cal G}^{n_i,h}_{n_f,s} \, \bigg \{ D_1+D_2+D_3 \bigg \} \, \rho^{\rm I}_{{\rm A},sh}(t)  
\end{eqnarray}
with
\begin{eqnarray}\label{ACoeffs}
 && D_1 = \sum_{\bf q} V^2_{\bf q}\, | F_{ss}({\bf q})- F_{hh}({-\bf q})|^2\, \gamma_r(\mathbf{q},0) , \\
 && D_2 = \sum_{\bf q} \big\{k_{hs}(\mathbf{q},\omega_{hs})  + k_{sh}(\mathbf{q},\omega_{sh}) \big\} , \\
 && D_3 = \sum_{n \neq s,h} \sum_{{\bf q}} \big\{ k_{sn}(\mathbf{q},\omega_{sn}) + k_{hn}(\mathbf{q},\omega_{hn}) \big\} .
\end{eqnarray}
This dissipator, describing the decoherence of the nondiagonal elements of the wave packet,
has three different contributions. $D_1$ originates from the vertex correction and contributes only
beyond the dipol approximation. $D_3$ represents the contributions of all intermediate transitions.
The transition between the contributing initial state $s$ and final state $h$ is hidden in $D_2$ and from which 
the transition rates for the wave packet can be defined
(see also Eq{~\eqref{A2coeff}} in App.{~\ref{App::RWA}} and the discussion there):
\begin{eqnarray}
 {\cal W}_{n_i \rightarrow n_f} && =  \sum_{s,h} {\cal G}^{n_i,h}_{n_f,s}\cdot w_{h \rightarrow s} 
\end{eqnarray}
with the atomic transition rate given in Eq.{~\eqref{transrate}}. 

In collision theory, the T-matrix $\hat T = \hat V + \hat V \hat G_0 \hat V + \hat V \hat G_0 \hat V \hat G_0 \hat V + \dots$
is used to calculate the cross sections and the transition rates. Comparison with the Born approximation implemented 
in the derivation of the QME{~\eqref{eq:bamastershmkom}} shows that only the first term $\hat V$ in T-matrix is taken into 
account. In order to obtain a better description of the collision effects in plasma, higher-order terms should be evaluated. 
We use a semiclassical approximation reported in Ref.{~\cite{BL9598}} to describe the modification of the transition rate 
due to the collision effects in plasma, which is given by
\begin{eqnarray}
 f(n, \varDelta n,\Theta) 
 = \ln \left[1+ \frac{1}{\Delta n \Theta (1+2.5 n \Theta /\Delta n)}\right]
 \cdot \left[\ln \left(1+ \frac{1}{\Delta n \Theta}\right)\right]^{-1},
 \quad \Theta = \sqrt{\frac{|E_n|}{k_BT}}
\end{eqnarray}
with $\Delta n = n-n'$ and the binding energy $E_n$ for the hydrogen atom. Therefore the modified 
transition rate for the wave packet description may be written as
\begin{equation}\label{Transitionrate}
 {\cal W}_{n_i \rightarrow n_f}  =  \sum_{s,h} {\cal G}^{n_i,h}_{n_f,s} \cdot w_{h \rightarrow s} \cdot f(h, |h-s|,\Theta) .
\end{equation}

In Fig.{~\ref{tranrate1}} we show the transition rates calculated from 
the expression{~\eqref{Transitionrate}} for two different values for the width of the hydrogenic
wave packet. Comparing with the experimental data of Helium, it can be seen that the transition rates calculated
with the wave packet width $\sigma_{n_0} = 0.75$, evaluated using Eq.{~\eqref{opwavepacket}} for the given plasma 
parameters $T$ and $n_{\rm pl}$, are in best agreement. 
The agreement reveals the coherent wave packet character of the Rydberg electron.
\begin{figure}[ht] 
  \centering
\includegraphics[width=0.45\textwidth]{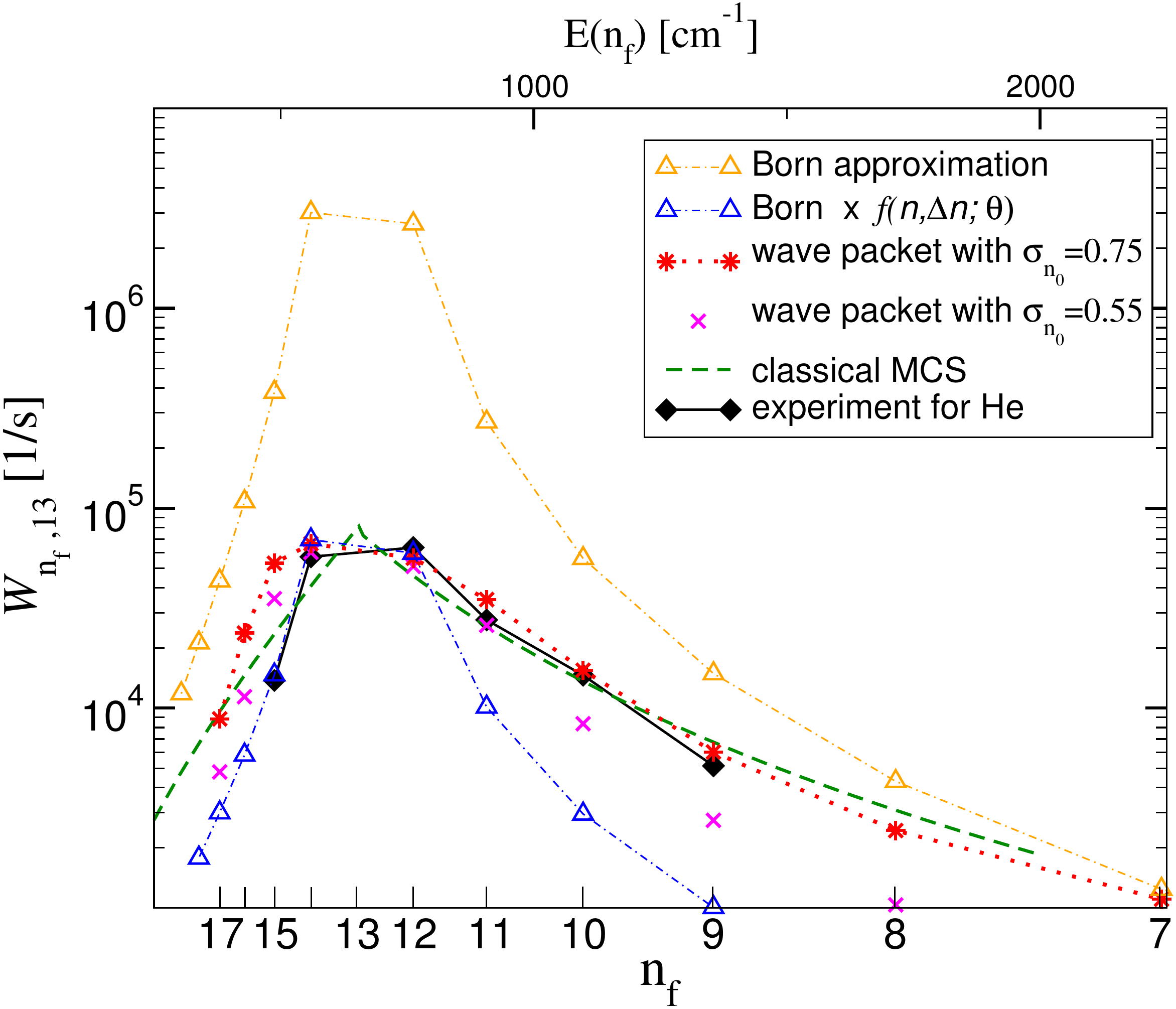}
  \caption{Transition rates of $n_i=13$ to near $n_f$ states induced at a $T=300$ K
  electron plasma with density $n_\mathrm{pl} = 10^9 \mathrm{cm}^{-3}$.
  calculated from
  the wave packet description with the width
  $\sigma_{n_0} = 0.55$ and $\sigma_{n_0} =0.75 $ compared to the results from classical Monte-Carlo 
  simulation (MCS){~\cite{VS80}}, the calculation in Born approximation with and without collision effects
  from Ref.{~\cite{GR06}} and experimental data{~\cite{DBD79}}.}
  \label{tranrate1}
\end{figure}

The comparison between the results of the classical Monte-Carlo simulation and the experimental data
indicates that a classical treatment is more appropriate to calculate the transition rates of the highly
excited states. In the classical Monte-Carlo simulation, the highly excited free
electron is treated as a point in an 18 dimensional phase space which behaves in
accordance with classical laws under the influence of the Coulomb interactions{~\cite{MK69}}.
From the quantum mechanical point of view, this treatment is equivalent to represent
the electron as an incoherent wave packet with vanishing width.  

Another comparison for the transition rates with the initial principal quantum number $n_i=40$
is shown in Fig.{~\ref{tranrate2}}.  
From the figure the validity of the wave packet description can be also verified from the agreement
between the results of classical Monte-Carlo simulations and the results calculated with the wave
packet width $\sigma_{n_0} = 2$.

\begin{figure}[ht]
 \centering
\includegraphics[width=0.45\textwidth]{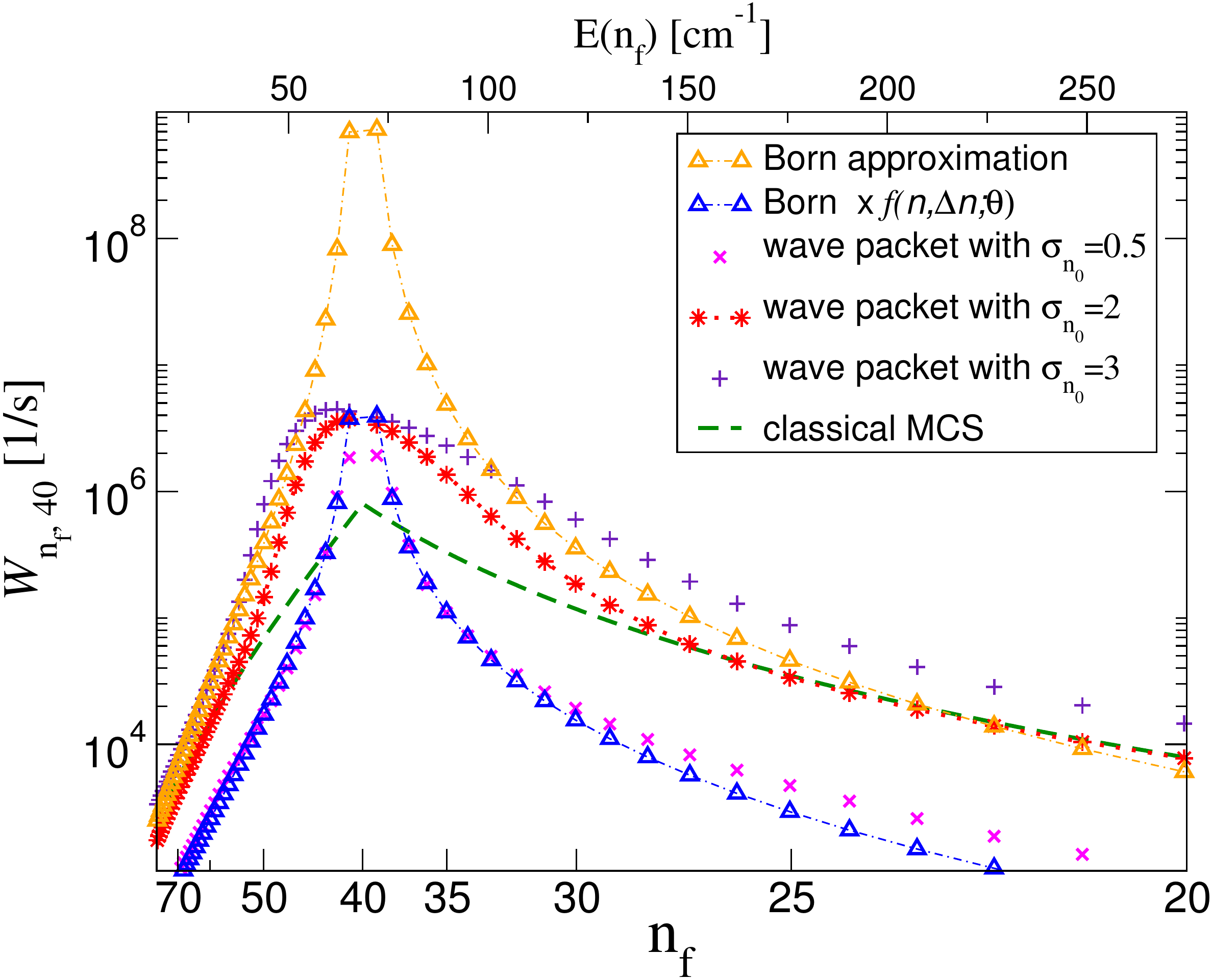}
 \caption{Transition rates of $n_i=40$ to near $n_f$ states induced by a $T=20$ K
 electron plasma with density $n_\mathrm{pl} = 10^9 \mathrm{cm}^{-3}$ calculated from 
 the wave packet description with the width
 $\sigma_{n_0} = 0.5, 2, 3$ compared to the results from classical Monte-Carlo 
 simulation{~\cite{VS80}} (green line)
 and the results in Born approximation with and without collision effects from Ref.{~\cite{BL9598}}.}
 \label{tranrate2}
\end{figure}

\section{Discussion and Conclusion}
\label{sec:Discussion}

We derived quantum master equations for an atom interacting with the charged particles of a plasma environment.
In Born-Markov approximation,
the influence function of the plasma environment is determined by the dynamical structure factor of the plasma.
As a consequence of the atom-plasma interaction, the electrons in highly excited Rydberg states become localized.
Localization of free electrons due to interactions with the environment is known from the quantum 
Brownian motion {\cite{JZKGKS03}}. 
This may be a good approximation in the limit of highly excited Rydberg states where the mean free path of the
electrons is small compared to the radius of the Kepler orbit. We derived a localization rate for electrons 
moving on a Kepler orbit, where the diagonal atomic formfactor appears.

Robust states are introduced as optimized wave packets.
The quantum diffluence of the wave packets is nearly compensated by the 
localization due to collisions with the surrounding plasma.
A critical quantum number $n_{\rm cr}$ is 
found. States with a lower quantum number are described by pure quantum states as solution of the 
atomic (hydrogen) Hamiltonian.  
For higher quantum numbers $n_0>n_{\rm cr}$, the superposition of different quantum states leads to 
a wave packet characterized by an average quantum number $n_0$.
Consequently, classical motion (Kepler ellipses) with a
corresponding Kepler radius $r_{\rm cl}= n_0^2 a_{\rm B}$ and an average azimuthal angle 
$\phi_{\rm cl}(t)$ is observed.
By construction, we are restricted to circular motion only. 
By avoiding the restriction to the 
Hilbert subspace of circular orbits ($l=m=n-1$),
more general Kepler orbits can be obtained taking into account all bound states for constructing the wave packet.

As another example for the use of the atomic master equation, the spectral line shape for transitions 
at low quantum numbers has been derived. The equivalence with a quantum statistical 
approach to profiles of spectral lines{~\cite{Guen95}} has been shown. 
After decoupling the ion and electron subsystems of the plasma environment, 
only the electron contribution to the spectral line shape 
has been considered (impact approximation). 
The standard description of the interaction with the plasma ions is 
the ionic microfield. 
The ionic structure factor determines the microfield distribution, and a superposition of the 
Stark shift in the ionic microfield and the electron contribution in impact approximation leads to the 
line profiles as derived from the unified theory \cite{Guen95}.

For comparison, the influence of the plasma ions on the line profile can also 
be calculated in Born approximation, similar to the treatment of the plasma electrons
using the impact approximation. The ions localize more strongly in comparison to the electrons
as a consequence of the larger localization rate (\ref{localrate}) if the electron mass 
is replaced by the ion mass. In other words, for electrons in plasma under
normal conditions the quantum description is applicable, whereas for ions the classical description
is more appropriate. The domains within the plasma density-temperature diagram, 
where the robust states of the ions are localized so that the concept of the classical ionic 
microfield can be justified while electrons should be treated quantum mechanically, have been outlined, for instance,
in Ref. \cite{GR07}. 

In the present work, we have shown that for electrons in Rydberg states localization may occur
owing to the interaction with the plasma environment.
Transition rates 
were calculated using robust quantum states formed by wave packets. 
Comparing with experiments and MCS results, the use of robust 
quantum states gives a better agreement with measured data and classical calculations than the 
approach using pure hydrogen eigenenergy states. Thus, the wave packet description which accounts for localization 
is more appropriate not only for the ions but also 
for the  electrons when considering highly excited Rydberg states. 
We performed exploratory calculations using the Brown circular-orbit wave packets. The approach can be 
further worked out to more general wave packets formed by the entire Hilbert space of the atomic states.

The existence of the plasma environment leads also to a reduction of the bound electron binding energy
because of the screening effects in the plasma. Consequently, bound states shift into the continuum which
is the so-called lowering  of the continuum edge \cite{KSK05}. 
This means there is a maximum principal quantum number $n_{\rm max}(T,n)$ for the 
Rydberg states at a given plasma temperature $T$ and a given density $n$ 
where separate bound states below the continuum can be identified. 
The reduction of the principal quantum number $n_{\rm cr}$, below which a pure hydrogen quantum state is robust,
as shown in Fig.{~\ref{boundary}}, has to be compared to
the pressure ionization of the plasma. Using the standard expression for the lowering of the ionization potential
in Ref.{~\cite{Salz98}}, estimations for the maximum principal quantum number $n_{\rm max}$ can be made.
For instance, at the plasma densities $10^9\, {\rm cm^{-3}}$ and $10^{15}\, {\rm cm^{-3}}$ 
the maximum principal quantum numbers are $n_{\rm max} \approx 200$ and  $n_{\rm max} \approx 20$, respectively.
Near the continuum edge, it is difficult to distinguish between the real continuum
edge and the point at which the spectroscopic series merges into a continuum 
due to line broadening. It would be of interest to investigate whether a wave packet description 
might be more suitable near the continuum edge. For this, the definition of the wave packet{~\eqref{wpkt}} 
should be extended to include the continuum states, similar to the case of free electrons where a 
Gaussian wave packet can be formed by plane wave states.

A fundamental issue in the theory of open quantum systems is that the subdivision of the total system into the 
reduced system and the bath is arbitrary and can be changed. Degrees of freedom of the bath
which are strongly coupled to the reduced system may be incorporated into the reduced system,
so that the bath contains only weakly coupled degrees of freedom which may be treated 
in Born-Markov approximation. Various approximations, in particular the Born-Markov approximation
and the rotating-wave approximation, performed in the present work can be improved in future work, 
see also Ref.~\cite{ZMR97}.
Furthermore, the electron in atom and the plasma electrons must be antisymmetrized so that 
exchange terms will occur. With respect to radiation processes, it is in general not the 
single electron which emits radiation but the whole reduced system which couples to the  radiation 
field. As an interesting application of this aspect, the treatment of radiation from many-electron
atoms, for instance the K$_\alpha$ radiation, is presently under investigation.

A main advantage of the QME for hydrogen Rydberg atom surrounded by a plasma 
is the use of robust states instead the
pure hydrogen eigenenergy states. The treatment of localization allows the transition to classical 
physics and the very efficient use of classical descriptions, for instance in molecular-dynamical simulations.
On the other side, QMEs are an essential ingredient to formulate a nonequilibrium 
approach for many-body systems, which can also be done on a very fundamental level as QED. The Rydberg atoms
considered in the present work are an interesting object to describe the transition from the quantum
microworld to macroscopic classical world where new properties such as trajectories emerge. \\

\begin{acknowledgments}
The authors acknowledge support within the DFG funded Special Research Unit SFB 652.  
\end{acknowledgments}


\appendix


\section{
Dynamical Structure Factor and Response Function}
\label{Kramers-Kronig}

The response function $\gamma_r$ is the real part of the Laplace transform of the density-density correlation function.
With the eigenstates $|\phi_n\rangle$ of the bath, $(\hat H_{\rm B}-\sum_c\mu_c \hat N_c)|\phi_n\rangle=B_n |\phi_n\rangle$,  
the spectral density of the density-density correlation function follows as
\begin{equation}
 I(\mathbf{q},\omega)=\frac{1}{e^2_e}\sum_{n,m}\frac{e^{-\beta B_n}}{\sum_{n'}e^{-\beta B_{n'}}} \langle \phi_n |\hat \varrho_{-{\bf q},{\rm B}}|\phi_m\rangle
  \langle \phi_m |\hat \varrho_{{\bf q},{\rm B}}|\phi_n\rangle 2 \pi \, \delta(\omega -B_n/\hbar+B_m/\hbar).
\end{equation}
The spectral density is the Fourier transform of the density autocorrelation function,
\begin{equation}
 \langle \hat \varrho_{-{\bf q},{\rm B}}(\tau)\hat \varrho_{{\bf q},{\rm B}}(0) \rangle_{\rm B} = e^2_e \, \int_{-\infty}^\infty \frac{d \omega}{2 \pi}
 I(\mathbf{q},\omega) e^{i \omega \tau} .
\end{equation}
We find
\begin{eqnarray}
 \Gamma_r(\mathbf{q},\omega)
 =\frac{e^2_e}{2\hbar^2}I(\mathbf{q},-\omega) + i {\cal P}\frac{e^2_e}{\hbar^2} \int_{-\infty}^\infty \frac{d \omega'}{2 \pi} 
 I(\mathbf{q},- \omega')\frac{1}{\omega - \omega'},
\end{eqnarray}
where $\cal P$ denotes the principal value of the integral.

Now we can use the fluctuation-dissipation theorem
\begin{equation} \label{verknuepfung}
 \gamma_r(\mathbf{q},\omega)=\frac{e^2_e}{\hbar^2}I(\mathbf{q},-\omega)
\end{equation}
and have for $S_r(\mathbf{q},\omega)$, which determines the Lamb shift, the Kramers-Kronig relation
\begin{equation}
 S_r(\mathbf{q},\omega)={\cal P} \frac{e^2_e}{\hbar^2}  \int_{-\infty}^\infty \frac{d \omega'}{2 \pi} 
 I(\mathbf{q},- \omega')\frac{1}{\omega-\omega'}={\cal P} \int_{-\infty}^\infty \frac{d \omega'}{2 \pi} 
2 \gamma_r(\mathbf{q},\omega')\frac{1}{\omega-\omega'}.
\end{equation}

The response function can be related to the dynamical structure factor (DSF) of the bath
which is defined via the Fourier
transform of the correlation function of the density fluctuation{~\cite{HM86}}:
\begin{equation}
 S_{\rm B}( \mathbf{q},\omega) = \frac{1}{2\pi n_{\rm pl} \Omega_0 e^2_e} \int^{\infty}_{-\infty} d\tau \  \mathrm{e}^{i \omega \tau} 
 \langle \delta \hat \varrho_{{\bf q},{\rm B}}(\tau) \delta \hat \varrho_{-{\bf q},{\rm B}}(0) \rangle_B,
\end{equation}
where $n_{\rm pl}$ is the electron density in plasma and 
$ \delta \hat \varrho_{{\bf q},{\rm B}}(\tau) = \hat \varrho_{{\bf q},{\rm B}}(\tau) - \langle \hat \varrho_{{\bf q},{\rm B}}(\tau) \rangle_B$ 
is the density fluctuation of the electrons. Because of the plasma environment in equilibrium, the condition 
$\langle \hat \varrho_{{\bf q},{\rm B}}(\tau) \rangle_B = e_e n_{\rm pl} \delta_{{\bf q},0}$ holds for all time. 
Then the DSF can be rewritten as 
\begin{eqnarray}
S_{\rm B}( \mathbf{q},\omega)  = \frac{1}{2\pi n_{\rm pl} \Omega_0 e^2_e} \int^{\infty}_{-\infty} d\tau \  \mathrm{e}^{i \omega \tau} 
 \langle \delta \hat \varrho_{{\bf q},{\rm B}}(\tau) \delta \hat \varrho_{-{\bf q},{\rm B}}(0) \rangle_B  
 = \frac{1}{2\pi n_{\rm pl} \Omega_0}  I(- \mathbf{q},\omega) + \frac{n_{\rm pl}}{2\pi  \Omega_0} \delta(\omega) \delta({\bf q}).
\end{eqnarray}
The last term in the above expression contributes only at $\omega = 0$ and ${\bf q}=0$. 
For dynamical processes this contribution can be neglected.

It can be obviously seen that the functions $\gamma_r(\mathbf{q},\omega), \ S_r(\mathbf{q},\omega),\  I(\mathbf{q},\omega)$ and 
$S_{\rm B}( \mathbf{q},\omega)$ are all related to the density-density correlation function and connected to each other, which means that
we need only one of them to construct the correlation function of the plasma environment. In this work we use the DSF $S_{\rm B}( \mathbf{q},\omega)$
which is directly related to the inverse dielectric function $\epsilon^{-1}(\mathbf{q},\omega)$ in plasma physics by employing the well-known
fluctuation-dissipation theorem{~\cite{KSK05}}:
\begin{equation}
 S_{\rm B}(\mathbf{q},\omega)  = \frac{\hbar}{\pi n_{\rm pl}} \frac{1}{\mathrm{e}^{\hbar \omega / k_B T }-1} \frac{\epsilon_0 q^2}{e_e^2} 
 \mathrm{Im}\left\{ \lim_{\delta \rightarrow 0^+} \epsilon^{-1}(\mathbf{q},\omega + i \delta)\right\}.
\end{equation}
The dielectric function can be treated by perturbation 
theory or numerical simulations as a quantum many-body problem.
An analytical approach calculating the dielectric function in
context of the linear response theory and the random
phase approximation can be found, e.g., in Refs.{~\cite{KSK05,GR13}}.

\section{Rotating Wave Approximation}
\label{App::RWA}

In this appendix we will investigate the influence of the RWA on the dynamics of the reduced system. The neglect of 
quickly oscillating terms in Eq. (\ref{paulibefore}) modifies the dynamics of the system. This procedure depends on the 
choice of the basis $|\psi_n \rangle$ which defines the diagonal and non-diagonal elements of the density matrix.

In contrast to the expressions given in subsec.{~\ref{sec:geqme:pauli}}, we here consider the result if performing the RWA in an 
earlier stage. The starting point is the QME{~\eqref{eq:bamastershmkom}} (interaction picture) with the influence function{~\eqref{influence7}}.
The RWA implies that the explicit dependence on $t-t_0$ disappears 
so that in Eq. (\ref{influence7}) only the 
terms with $m=n'$ and $m'=n$ contribute. We find
\begin{align}\label{influenceRWA2}
\hat {\cal D}^{\rm I}_{(1)}(t,t_0)  = - \sum_{nn',\bf{q}} K_{n'n;n'n}({\bf q},\omega_{n'n})
 \Big\{ \hat T_{n'n'} \hat \rho^{\rm I}_{\rm A}(t,t_0) -   \hat T_{nn'} \hat \rho^{\rm I}_{\rm A}(t,t_0) \hat T_{n'n} \Big\} + \rm {h.c.}
\end{align}
In addition  the explicit dependence on $t-t_0$ disappears for $n'=n$ and $m'=m$ so that
\begin{align}\label{influRWA2}
\hat {\cal D}^{\rm I}_{(2)}(t,t_0)  = - \sum_{mn,\bf{q}} K_{mm;nn}({\bf q},\omega_{mm})
 \Big\{ \hat T_{mm} \hat \rho^{\rm I}_{\rm A}(t,t_0) \delta_{mn} -   \hat T_{mm} \hat \rho^{\rm I}_{\rm A}(t,t_0) \hat T_{nn} \Big\} + \rm {h.c.} 
\end{align}
The term $m=n$ in the sum of $\hat {\cal D}^{\rm I}_{(2)}$ gives the same contribution as in $\hat {\cal D}^{\rm I}_{(1)}$
if $n'=n$. To avoid this double counting, the corresponding contributions in $\hat {\cal D}^{\rm I}_{(2)}$ should be substracted.
The correct contribution can be expressed as
\begin{align}\label{inflRWA4}
 \hat {\cal D}^{\rm I}_{(2)}(t,t_0)  & = \sum_{n'\neq n,\bf{q}} V_{\bf q}^2 \, F_{nn}({\bf q}) F^*_{n'n'}({\bf q})\,
 \Big( \Gamma_{r}(\mathbf{q},\omega_{n'n'} ) + \Gamma^*_{r}(\mathbf{q},\omega_{n'n'} ) \Big ) \cdot
 \hat T_{n'n'} \hat \rho^{\rm I}_{\rm A}(t,t_0) \hat T_{nn} \nonumber \\ 
 & = \sum_{n'\neq n,\bf{q}} V_{\bf q}^2 \, F_{nn}({\bf q}) F^*_{n'n'}({\bf q})\,
 \gamma_{r}(\mathbf{q},0 ) \cdot \hat T_{n'n'} \hat \rho^{\rm I}_{\rm A}(t,t_0) \hat T_{nn}.
\end{align}
In dipole approximation, this expression yields no contribution. Beyond dipole approximation this term contributes only
to the vertex correction. Alltogether, the influence function in RWA follows as 
\begin{equation}
 \hat{\cal D}^{\rm I}_{\rm RWA}(t,t_0)=\hat {\cal D}^{\rm I}_{(1)}(t,t_0)
+ \hat {\cal D}^{\rm I}_{(2)}(t,t_0).
\end{equation}
The influence function $\hat{\cal D}^{\rm I}_{(1)}(t,t_0)$ can be transformed into a more transparent form.
With the decomposition of the response function $\Gamma_r( \mathbf{k},\omega)${~\eqref{realGamma}},
the influence function {\eqref{influenceRWA2}} and {\eqref{influRWA2}} can be rewritten as 
\begin{eqnarray} \label{influenceRWA3}
\hat {\cal D}^{\rm I}_{(1)}(t,t_0) & =& - \sum_{nn',\bf{q}} 
 \bigg\{ \frac{1}{2} k_{n'n}({\bf q},\omega_{n'n}) \Big[ \big \{ \hat T_{n'n'} \hat \rho^{\rm I}_{\rm A}(t,t_0)
 + \hat \rho^{\rm I}_{\rm A}(t,t_0) \hat T_{n'n'} \big\}
 - 2\, \hat T_{nn'} \hat \rho^{\rm I}_{\rm A}(t,t_0) \hat T_{n'n} \Big] \nonumber \\ && 
 - i  \sum_{nn',\bf{q}} V_{\bf q}^2 F_{n'n}({\bf q}) F_{nn'}({-\bf q})\, S_{r}(\mathbf{q},\omega_{n'n} ) 
 \Big[ \hat T_{n'n'} \hat \rho^{\rm I}_{\rm A}(t,t_0) - \hat \rho^{\rm I}_{\rm A}(t,t_0) \hat T_{n'n'}  \Big]
 \bigg\}.
\end{eqnarray}
The last term in Eq. (\ref{influenceRWA3}) can be rewritten as commutator describing the reversible Hamiltonian dynamics
which in fact represents the line shift of the eigenenergy levels of the atomic system induced by the coupling 
to the background as known from the coupling to the radiation field. The terms in the first line of the influence 
function (\ref{influenceRWA3}) are responsible for the transition processes of atoms. 
Since $F_{nn}({\bf q}) F^*_{n'n'}({\bf q})$ is a complex quantity,
the influence function $\hat{\cal D}^{\rm I}_{(2)}(t,t_0)$ can be also decomposed into a real part 
\begin{align} \label{realD2}
 \hat{\rm D}_{(2)}[\hat \rho_{\rm A}(t)] = 
 \sum_{n'\neq n,\bf{q}} V_{\bf q}^2 \cdot {\rm Re} \Big\{ F_{nn}({\bf q}) F^*_{n'n'}({\bf q}) \Big\} \cdot 
 \gamma_{r}(\mathbf{q},0 ) \cdot \hat T_{n'n'} \hat \rho^{\rm I}_{\rm A}(t,t_0) \hat T_{nn}
\end{align}
and an imaginary part 
\begin{align}\label{imag2}
  \hat{H}^{(2)}_{\mathrm{shift}}  = 
   \sum_{n'\neq n,\bf{q}} V_{\bf q}^2 \cdot {\rm Im} \Big\{ F_{nn}({\bf q}) F^*_{n'n'}({\bf q}) \Big\} \cdot 
 \gamma_{r}(\mathbf{q},0 ) \cdot \hat T_{n'n'} \hat \rho^{\rm I}_{\rm A}(t,t_0) \hat T_{nn}.
\end{align}

We go back to the Schr{\"o}dinger picture with Eq. (\ref{IP}), 
$\hat \rho^{\rm I}_{\rm A}(t,t_0)= {\rm e}^{i (\hat H_{\rm A} + \hat H_{\rm B} ) (t-t_0) /\hbar} \,  
\hat \rho_{\rm A}(t)\,  {\rm e}^{-i (\hat H_{\rm A} + \hat H_{\rm B} ) (t-t_0) /\hbar} $.
Then the atomic QME becomes
\begin{eqnarray} \label{qmeRWA}
&& \frac{\partial \hat \rho_{\rm A}(t)}{\partial t} - 
  \frac{1}{i\hbar}\left[\hat{H}_{\rm A}+\hat{H}^{(1)}_{\mathrm{shift}}+\hat{H}^{(2)}_{\mathrm{shift}}, 
  \hat \rho_{\rm A}(t) \right] 
= {\hat {\rm D}}_{(1)}[\hat \rho_{\rm A}(t)] +  \hat {\rm D}_{(2)}[\hat \rho_{\rm A}(t)] 
\end{eqnarray}
with the shift Hamiltonian operator 
\begin{align} \label{imag1}
\hat{H}^{(1)}_{\mathrm{shift}}  = \sum_{nn',\bf{q}} V_{\bf q}^2 | F_{n'n}({\bf q}) |^2
S_{r}(\mathbf{q},\omega_{n'n} ) \hat T_{n'n'} , 
\end{align}
which is related the shift of the eigenenergies.
The dissipator $\hat{ \mathrm{D}}_{(1)}[\hat \rho_{\rm A}(t)]$, 
which is the real part of the influence function is given in Schr{\"o}dinger picture by
\begin{eqnarray}
\label{influenceRWA7}
{\hat {\rm D}}_{(1)}[\hat \rho_{\rm A}(t)] = \sum_{nn',\bf{q}} k_{n'n}(\mathbf{q},\omega_{n'n} ) \,\,
 \left[ \hat T_{nn'} \hat \rho_{\rm A}(t) \hat T_{n'n} -\frac{1}{2}\left\{\hat T_{n'n'} , \hat \rho_{\rm A}(t)
 \right\}_{+} \right],
\end{eqnarray}
where the curly brackets denote the anticommutator. Without the contributions from
$\hat {\cal D}^{\rm I}_{(2)}[\hat \rho_{\rm A}(t)]$,
the QME{~\eqref{qmeRWA}} has the Lindblad form. 
Generally, by performing the RWA here we can render the QME in the Lindblad form 
in which the terms describing atomic emissions and absorptions can be seperated as shown in Ref.{~\cite{BreuerPetruccione}}.
However, we should point out that the neglecting of the term $\hat {\cal D}^{\rm I}_{(2)}[\hat \rho_{\rm A}(t)]$ yields
an incorrect description of the dissipative system beyond dipole approximation.


We implement the matrix representation of the QME (\ref{qmeRWA}) in the Schr{\"o}dinger picture with Eq. (\ref{IP}), 
then the atomic QME in RWA becomes
\begin{align}
& \frac{\partial \rho_{{\rm A},if}(t)}{\partial t}+i \omega_{if}\rho_{{\rm A},if}(t) \\
= & - \sum_{n,\bf{q}} \Big[ K_{in;in}({\bf q},\omega_{in}) + K^*_{fn;fn}({\bf q},\omega_{fn}) \Big] \cdot \rho_{{\rm A},if}(t) \nonumber \\
&  + \delta_{if} \sum_{n,\bf{q}} \Big[ K_{ni;ni}({\bf q},\omega_{ni}) + K^*_{ni;ni}({\bf q},\omega_{ni}) \Big] \cdot \rho_{{\rm A},nn}(t) \\
& + (1-\delta_{if}) \sum_{\bf{q}}\Big[ K_{ii;ff}({\bf q},\omega_{ii}) + K^*_{ff;ii}({\bf q},\omega_{ff}) \Big] \cdot \rho_{{\rm A},if}(t).
\end{align}
The last contribution comes from $ \hat{\cal D}^{\rm I}_{(2)}(t,t_0) $, Eq. (\ref{inflRWA4}).

On the other hand, we can also study the dissipator {\eqref{influenceRWA7}} in its matrix representation. 
The Pauli equation resulting from the diagonal matrix elements of the the dissipator {\eqref{influenceRWA7}} 
is given by
\begin{eqnarray} \label{PauliRWA1}
 \frac{\partial P^{(1)}_i(t)}{\partial t}  
 = \sum_{n,{\bf q}}  \left \{  k_{ni}(\mathbf{q},\omega_{ni}) P^{(1)}_n(t) 
 - k_{in}(\mathbf{q},\omega_{in}) P^{(1)}_i(t)  \right \} .
\end{eqnarray}
This relation coincides with the Pauli equation (\ref{PauliEqu}) because the contribution
$\hat{\cal D}^{\rm I}_{(2)}(t,t_0)$
does not affect the behavior of the population numbers given by the diagonal terms of the density matrix. 
Note that in comparison to the derivation given in this appendix two additional terms occur 
in Eq.{~\eqref{paulibefore}}, which contain nondiagonal 
matrix elements $\rho_{{\rm A},if}(t)$.
The neglecting of these additional terms is only valid if the differences of neighbored eigenenergies $E_n$ of
the basis $| \psi_n \rangle$ are enough large so that these terms oscillate quite quickly. 
In the case of Rydberg states,
these terms oscillating with  
frequency $\omega_{if}$ are also relevant and cannot be ignored any more.

The nondiagonal matrix elements of the dissipator {\eqref{qmeRWA}}, i.e. ${\hat{\rm D}}_{\rm RWA}[\hat \rho_{\rm A}(t)] 
= {\hat {\rm D}}_{(1)}[\hat \rho_{\rm A}(t)] +  \hat {\rm D}_{(2)}[\hat \rho_{\rm A}(t)]$, 
can be represented as 
\begin{eqnarray} \label{transitionRWA2}
 \frac{\partial \rho_{{\rm A},if}(t)}{\partial t}+i {\tilde \omega}_{if}\, \rho_{{\rm A},if}(t)
 = \langle \psi_{i}| {\hat {\rm D}}_{\rm RWA}[\hat \rho_{\rm A}(t)] |\psi_{f} \rangle 
 = - \frac{1}{2} \bigg \{ d_1+d_2+d_3 \bigg \} \, \rho_{{\rm A},if}(t) 
\end{eqnarray}
with the modified transition frequency ${\tilde \omega}_{if}$ due to the shift Hamiltonian in Eq.{~\eqref{qmeRWA}}.
The contributions $d_1,\ d_2$ and $d_3$ are defined similarly as in Eq.{~\eqref{ACoeffs}},
\begin{eqnarray}
 && d_1 = \sum_{\bf q} V^2_{\bf q}\, | F_{ii}({\bf q})- F_{ff}({-\bf q})|^2\, \gamma_r(\mathbf{q},0) , \\
 && d_2 = \sum_{\bf q} \big\{ k_{if}(\mathbf{q},\omega_{if}) + k_{fi}(\mathbf{q},\omega_{fi}) \big\} , \label{A2coeff} \\ 
 && d_3 = \sum_{n \neq i,f} \sum_{{\bf q}} \big\{ k_{in}(\mathbf{q},\omega_{in}) + k_{fn}(\mathbf{q},\omega_{fn}) \big\} .
\end{eqnarray}
The mixed contribution in $d_1$ originates from the dissipator 
$\hat {\rm D}_{(2)}[\hat \rho_{\rm A}(t)]${~\eqref{realD2}}, whereas another two contributions 
belong to the dissipator $\hat {\rm D}_{(1)}[\hat \rho_{\rm A}(t)]${~\eqref{influenceRWA7}}.
It can be seen that the expression{~\eqref{A2coeff}} relates directly to the transition rates of the atomic eigenstates comparing with
the Pauli equation{~\eqref{PauliEqu}} for a given two-levels system transition, which gives a clue to define the transition rates 
for the Rydberg wave packet via the QME as explained in Sec.{~\ref{wp:transrate}}.

For the sake of investigating the effect of the RWA we return to the atomic QME{~\eqref{influence10}} which reads in 
the Schr{\"o}dinger picture
\begin{equation} \label{influenceRWA6}
 \frac{\partial \rho_{{\rm A},if}(t)}{\partial t}+i \omega_{if}\rho_{{\rm A},if}(t) 
 = \langle \psi_{i} | \hat{\cal D}[\hat \rho_{\rm A}(t)] | \psi_{f} \rangle
\end{equation}
with the influence function [remember $\rho^{\rm I}_{{\rm A},mn}(t,t_0)= \e^{i \omega_{mn} (t-t_0)} \rho_{{\rm A},mn}(t)$]

\begin{eqnarray}\label{matrixelementRWA}
 \langle \psi_{i} |\hat {\cal D}[\hat \rho_{\rm A}(t)] | \psi_{f} \rangle &
 = &- \sum_{mn,{\bf q}} \Big \{  K_{mn;in}({\bf q},\omega_{mn}) \, \rho_{{\rm A},mf}(t)
 +\, K^{*}_{mn;fn}({\bf q},\omega_{mn}) \, \rho_{{\rm A},im}(t) \nonumber \\&& \qquad 
 -\, \big[ K_{mi;fn}({\bf q},\omega_{mi}) 
+ K^{*}_{nf;mi}({\bf q},\omega_{nf}) \big ] 
 \, \rho_{{\rm A},mn}(t) \Big \} .
\end{eqnarray}

The RWA for the non-diagonal terms means we should set $m=i$ in the first term, $m=f$ in the second term and $m=i,\ n=f$ in the third term
of the influence function {\eqref{matrixelementRWA}}. By using the decomposition of the complex response function 
$\Gamma_r(\mathbf{q},\omega) =\gamma_r(\mathbf{q},\omega)/2 + i S_r (\mathbf{q},\omega) $ we obtain 
the same expression as Eq.{~\eqref{transitionRWA2}}.

In principle, the RWA by the removal of terms that oscillate
quickly with respect to some characteristic time scales of the
system yields is problematic as pointed out by
different authors. It depends on the choice of the basis $| \psi_n \rangle$ for the representation of the density matrix, 
and in the case of small energy differences of neighbored eigenenergies $E_n$ the oscillation may become not quick enough 
compared to the characteristic time scales of the
system. Also recently the RWA is under discussion. In a study of the spontaneous emission of a two-level system, 
Agarwal found that the RWA gives an incorrect value for environmentally induced frequency shifts with respect to the system frequency {\cite{Agar7173}}.
Fleischhauer studied the photodetection without the RWA, finding that for ultrashort pulses, whose 
length is of the order of the oscillation period, the mean number of photocounts with the RWA and without the RWA are substantially different {\cite{Flei98}}.
Recently, Fleming {\etal} investigated the validity of the RWA in an open quantum system and argued that the quantum state resulting from the RWA
is inappropriate for calculating the detailed properties of the state dynamics such as entanglement dynamics {\cite{Fle10}}.
In Ref. {\cite{Maj13}}, Majenz {\etal} showed that the RWA leads to the missing of important qualitative features of the population dynamics in a special
three-level model. Recently, M\"akel\"a and M\"ott\"onen {\cite{Mak13}} discovered that the RWA yields an impressive reduction in the amount of non-Markovianity 
and is problematic if non-Markovian dynamics is of essential relevance.

In this work, we found that the RWA performed in Eq. (\ref{influence7}) be neglecting
$\hat {\cal D}^{\rm I}_{(2)}(t,t_0)$ leads to a QME in Lindblad form. However, 
the term $\hat {\cal D}^{\rm I}_{(2)}(t,t_0)$ has a significant contribution in some special cases, for example,
the vertex correction of the spectral line profiles.
On the other hand, if the RWA is performed in the matrix representation, the contribution of 
$\hat {\cal D}^{\rm I}_{(2)}(t,t_0)$ can be automatically included in the influence function.
If the RWA is carried out prematurely, it will be inappropriate
to describe the dissipative properties of the relevant atomic system (Rydberg states) 
and result in erroneous transition rates.

\section{Derivation of the QME for the Spectral Profile }
\label{App::Profile}

As mentioned before, the charge density operator in the case of spectral line profiles is given by 
\begin{eqnarray}\label{chargeDensity}
  \hat \varrho^{\rm I}_{{\bf q}, {\rm S}} (\omega) 
 = \sum_{n'>n} e_e\, \hat T_{n'n}^{-}  \cdot F_{n'n} ({\bf q})\,  \delta(\omega - \triangle \omega_{n'n})
 - \sum_{n'<n} e_e\, \hat T_{n'n}^{+} \cdot F_{n'n} ({\bf q}) \, \delta(\omega +\triangle \omega_{nn'}) .
\end{eqnarray}
The first term in{~\eqref{chargeDensity}} describes the absorption process which the second one signifies the emission process.
Inserting this expression into the influence function{~\eqref{influence2}}, we obtain a new influence function 
including both emission and absorption terms, which can be used as the starting point to derive the spectral line profile. 
The terms representing the emission processes can be selected by using the matrix element representation
$\langle \tilde f | {\cal D}^{\rm I}[\hat \rho_S(t)] |\tilde i \rangle$ with the change of the photon number
$\triangle N = N_f(\bar {\bf k}) -N_i(\bar {\bf k}) = 1$:
\begin{equation}
 \langle \tilde f | {\cal D}^{\rm I}[\hat \rho^{\rm I}_S(t)] | \tilde i \rangle 
 = - A_1 - A_2 + A_3 + A_4
\end{equation}
with 
\begin{eqnarray}
  && A_1 = \sum_{n>f,m<n,{\bf q }} \exp[  i (-\triangle \omega_{nm} + \triangle \omega_{nf} ) t ] \,
     K_{fn;nm}(\mathbf{q},  \triangle \omega_{nm} )
     \, \langle \psi_m,\triangle N | \hat \rho_S(t) | \psi_i  \rangle \nonumber \\ && \qquad
     + \sum_{n<f,m>n,{\bf q }} \exp[  i (\triangle \omega_{mn} - \triangle \omega_{fn} ) t ] \, 
     K_{fn;nm}(\mathbf{q}, - \triangle \omega_{mn} ) 
      \, \langle \psi_m,\triangle N | \hat \rho_S(t) | \psi_i  \rangle ,\nonumber 
\end{eqnarray}
\begin{eqnarray}
 && A_2 =  \sum_{n>i,m<n,{\bf q }} \exp[ - i (-\triangle \omega_{nm} + \triangle \omega_{ni} ) t ] \,
     K^*_{mn;ni}(\mathbf{q},  \triangle \omega_{nm} )
      \, \langle \psi_f,\triangle N | \hat \rho_S(t) | \psi_m  \rangle \nonumber \\ && \qquad
     + \sum_{n<i,m>n,{\bf q }} \exp[ - i (\triangle \omega_{mn} - \triangle \omega_{in} ) t ] \,
     K^*_{mn;ni}(\mathbf{q},  \triangle \omega_{nm} ) 
      \, \langle \psi_f,\triangle N | \hat \rho_S(t) | \psi_m  \rangle ,\nonumber
\end{eqnarray}
\begin{equation}
  A_3 =  \sum_{i>n,m<f,{\bf q }} \exp[  i (-\triangle \omega_{fm} + \triangle \omega_{in} ) t ] \,
      \big \{  K_{ni;fm}(\mathbf{q},  \triangle \omega_{fm} ) +  K^*_{mf;in}(\mathbf{q},  \triangle \omega_{in} ) \big \}
       \, \langle \psi_m,\triangle N | \hat \rho_S(t) | \psi_n  \rangle ,\nonumber
\end{equation}
\begin{equation}
 A_4 =  \sum_{i<n,m>f,{\bf q }} \exp[  i (-\triangle \omega_{fm} + \triangle \omega_{in} ) t ] \,
      \big \{   K_{ni;fm}(\mathbf{q}, - \triangle \omega_{mf} ) +  K^*_{mf;in}(\mathbf{q}, - \triangle \omega_{ni} ) \big \} 
       \, \langle \psi_m,\triangle N | \hat \rho_S(t) | \psi_n  \rangle ,\nonumber
\end{equation}
where the indexes $m$ and $n$ are interchanged.
These terms can be further simplified in RWA. This means that we can set $m=f$ in $A_1$,
$m=i$ in $A_2$, and $m=f, \ n = i$ in $A_3$ and $A_4$.
The QME in RWA becomes
\begin{eqnarray}
 \frac{\partial \rho_{{\rm S},fi}^{\rm I}(t)}{\partial t} = 
 -  \Gamma_{fi}^{{\rm BS}}(\omega_{\bar {\bf k}})  \rho_{{\rm S},fi}^{\rm I}(t)
 + \Gamma_{fi}^{\nu}\, \rho_{{\rm S},fi}^{\rm I}(t)
\end{eqnarray}
with a coefficient $\Gamma_{fi}^{BS}(\omega_{\bar {\bf k}}) $ describing the shift of the eigenenergy levels and the pressure broadening
\begin{eqnarray}
 \Gamma_{fi}^{{\rm BS}}(\omega_{\bar {\bf k}}) 
 = \sum_{n,{\bf q}} \big\{ K_{nf;fn}(\mathbf{q},  \triangle \omega_{nf}) + K_{nf;fn}(\mathbf{q}, - \triangle \omega_{fn})
 + K^{*}_{ni;in}(\mathbf{q},  \triangle \omega_{ni}) +  K^{*}_{ni;in}(\mathbf{q}, - \triangle \omega_{in})\big \}
\end{eqnarray}
and a coefficient $\Gamma_{fi}^{\rm V}$ describing the vertex correction
\begin{eqnarray}
  \Gamma_{fi}^{\rm V}
  = \sum_{{\bf q}} \big\{ K_{ii;ff}(\mathbf{q},  \triangle \omega_{ff}) + K_{ii;ff}(\mathbf{q}, - \triangle \omega_{ff}) 
  + K^{*}_{ff;ii}(\mathbf{q},  \triangle \omega_{ii}) +  K^{*}_{ff;ii}(\mathbf{q}, - \triangle \omega_{ii}) \big\} ,
\end{eqnarray}
which has no dependence on $\omega_{\bar {\bf k}}$ in our approximation.

\section{Collisional Decoherence of a Rydberg Electron in Plasma}
\label{colldecoh}

Following the method represented in the book of Joos {\etal}{~\cite{JZKGKS03}}
the reduced density matrix for the Rydberg electron can be derived under
the assumptions of recoil-free collisions and elastic scattering
\begin{eqnarray}
 \rho({\bf R}_{n}, {\bf R}^{'}_n) \rightarrow \rho({\bf R}_{n}, {\bf R}^{'}_n)
 \cdot \bigg\{ 1 + \sum_{\bf k'} \Big[ 1 - {\rm e}^{i({\bf k-k'}) \cdot ({\bf R}_{n} - {\bf R}^{'}_n) } \Big]
  |\langle {\bf k'}, n | \hat T | {\bf k}, n \rangle |^2 \bigg\}, 
\end{eqnarray}
where the T-matrix is given by $\hat T = \hat V + \hat V \hat G_0 \hat V + \hat V \hat G_0 \hat V \hat G_0 \hat V + ...$.
In the elastic scattering process the principal quantum number $n$ of the Rydberg electron does not change, this means, the Rydberg
electron motions along the classical Kepler orbit.
For the bound electrons the T-matrix in Born approximation can be represented as
\begin{eqnarray}
 \langle {\bf k'}, n | \hat T | {\bf k}, n \rangle = V_{\bf q} \, F_{nn}({\bf q}) \delta(E_{\bf k} - E_{\bf k'})
\end{eqnarray}
with ${\bf q} = {\bf k-k'}$ and $E_{\bf k} = {\bf k}^2/(2m_e)$. 
$V_{\bf q}$ denotes the interaction potential and $F_{nn}(\bf q)$ is diagonal atomic form factor.

In Born approximation we have
\begin{eqnarray}\label{SingleScatt}
 A && :=  \sum_{\bf k'} \Big[ 1 - {\rm e}^{i({\bf k-k'}) \cdot ({\bf R}_{n} - {\bf R}^{'}_n) } \Big]
  |\langle {\bf k'}, n | \hat T | {\bf k}, n \rangle |^2 \nonumber \\ && 
  = \frac{\Omega_0}{(2 \pi)^3} \int d^3{\bf k'} \, \Big[ 1 - {\rm e}^{i({\bf k-k'}) \cdot ({\bf R}_{n} - {\bf R}^{'}_n) } \Big] \,
   V^2_{\bf q} \, F^2_{nn}({\bf q}) \delta^2(E_{\bf k} - E_{\bf k'}) \nonumber \\
 && = \frac{\Omega_0 m_e T}{(2 \pi \hbar)^3 k} \int_{0}^{2k} d q\,  q V^2_{\bf q} \, F^2_{nn}({\bf q})
 \, \Big[ 1 - {\rm e}^{i{\bf q} \cdot ({\bf R}_{n} - {\bf R}^{'}_n) } \Big] .
\end{eqnarray}
In the third line the integrals over $k'$ and $\phi$ have been carried out and the integral over $\theta$ is replaced by
$\int_{|k-k'|}^{k+k'} d q$ by using the relation $q^2 = k^2 + k'^2 - 2kk'\cos \theta$.
The squared delta function is evaluated by using the Fourier representation of the delta function 
\begin{eqnarray}
 \delta^2(E_{\bf k} - E_{\bf k'}) = \delta(E_{\bf k} - E_{\bf k'}) \cdot
 \lim_{T \rightarrow \infty} \frac{1}{2 \pi \hbar} \int_{-T/2}^{T/2} dt \, e^{i(E_{\bf k} - E_{\bf k'})t/\hbar}
 = \frac{m_e}{\hbar^2 k'} \delta(k-k') \cdot \lim_{T \rightarrow \infty} \frac{T}{2 \pi \hbar}.
\end{eqnarray}


For a collection of $N$ independent scattering events in plasma, the above expression{~\eqref{SingleScatt}} should 
be multiplied by a factor $N$. For the momentum distribution $P({\bf q})$ of the plasma environment,
the classical Maxwell-Boltzmann distribution is taken.
where the momentum distribution $P(\hbar {\bf k})$ of the plasma environment, assumed to
fulfill Maxwell-Boltzmann distribution 
$P({\bf q}) = (\frac{\hbar^2}{2 \pi m_e k_B T})^{3/2} \exp[-\hbar^2 q^2/(2 m_e k_B T)]$ is taken.
We find
\begin{eqnarray}\label{Nscatt}
 A= \frac{N \Omega_0 T}{\pi (2 \pi \hbar)^2 } \sqrt{\frac{m_e}{2 \pi k_B T}} \int_{0}^{\infty} d q\,  q V^2_{\bf q} \, F^2_{nn}({\bf q})
 \,{\rm e}^{-\hbar^2 q^2/(8 m_e k_B T)}\, \Big[ 1 - {\rm e}^{i{\bf q} \cdot ({\bf R}_{n} - {\bf R}^{'}_n) } \Big].
\end{eqnarray}
For the scattering process described here we have the time evolution of the reduced density matrix (QME) by taking
the differential limit of small $T$
\begin{eqnarray}\label{motioneq1}
 \frac{\rho({\bf R}_{n}, {\bf R}^{'}_n,T)-\rho({\bf R}_{n}, {\bf R}^{'}_n,0)}{T} 
 = \frac{N \Omega_0 }{\pi (2 \pi \hbar)^2 } \sqrt{\frac{m_e}{2 \pi k_B T}} \int_{0}^{\infty} d q\,  q V^2_{\bf q} \, F^2_{nn}({\bf q})
 \,{\rm e}^{-\hbar^2 q^2/(8 m_e k_B T)}\, \Big[ 1 - {\rm e}^{i{\bf q} \cdot ({\bf R}_{n} - {\bf R}^{'}_n) } \Big].
\end{eqnarray}
To avoid the divergence of the integral in{~\eqref{motioneq1}}, the Debye potential {\cite{KSK05}} can be used. 
As the next step we can use the long-wavelength limit to evaluate {\eqref{motioneq1}}, i.e. we can expand
the exponential function
${\rm e}^{i{\bf q} \cdot ({\bf R}_{n} - {\bf R}^{'}_n) }$ up to second order and obtain
the QME in the long-wavelength limit
\begin{equation}
 \frac{\partial}{\partial t}  \rho({\bf R}_{n}, {\bf R}^{'}_n,t) 
 = - \frac{N \Omega_0 }{\pi (2 \pi \hbar)^2 } \sqrt{\frac{m_e}{2 \pi k_B T}} \int_{0}^{\infty} d q\,  q V^2_{\bf q} \, F^2_{nn}({\bf q})
 \,{\rm e}^{-\hbar^2 q^2/(8 m_e k_B T)}\,( {\bf q} \cdot ({\bf R}_{n} - {\bf R}^{'}_n) )^2 .
\end{equation}

As shown in Sec.{~\ref{wp:robust}}, the Rydberg electron moves along the Kepler orbit, i.e. 
${\bf R}_{n} = (n^2 a_{\rm B}, \pi/2, \phi)$ and ${\bf R}^{'}_{n} = (n^2 a_{\rm B}, \pi/2, \phi')$. This assumption allows
us to calculate the term $( {\bf q} \cdot ({\bf R}_{n} - {\bf R}^{'}_n) )^2$ by averaging it over all possible
directions $({\bf R}_{n} - {\bf R}^{'}_n)$:
$( {\bf q} \cdot ({\bf R}_{n} - {\bf R}^{'}_n) )^2= q^2\cdot (x - x')^2/3$ with $x=r_{\rm cl} \phi$.
Then we have 
\begin{eqnarray}
 \frac{\partial}{\partial t}  \rho(x,x',t) 
 = - \Lambda_{R_n} \cdot  (x - x')^2
\end{eqnarray}
with the localization rate defined by
\begin{equation}\label{localratebound}
 \Lambda_{R_n} = \frac{N \Omega_0 }{\pi (2 \pi \hbar)^2 }  \int_{0}^{\infty} d q\, \frac{q^4}{3}  V^2_{\bf q} \, F^2_{nn}({\bf q})
 \,\sqrt{\frac{m_e}{2 \pi k_B T q^2} }\,{\rm e}^{-\hbar^2 q^2/(8 m_e k_B T)}.
\end{equation}
For a free electron moving in the plasma environment we can recover the localization rate from
the expression{~\eqref{localratebound}}
by setting $F^2_{nn}({\bf q}) = 1$, which coincides with the result reported 
in Ref.{~\cite{Vac01}} up to a factor $2 \pi^2$.

\section*{Bib}

\end{document}